\documentclass[lettersize,journal]{IEEEtran}
\usepackage{amsmath,amsfonts}
\usepackage{array}
\usepackage[caption=false,font=normalsize,labelfont=sf,textfont=sf]{subfig}
\usepackage{textcomp}
\usepackage{stfloats}
\usepackage{url}
\usepackage{verbatim}
\usepackage{graphicx}
\usepackage{cite}
\usepackage{amssymb,amsthm,autobreak,bm}
\usepackage[ruled,vlined]{algorithm2e}

\usepackage[english]{babel}
\addto\captionsenglish{%
  \renewcommand{\figurename}{Fig.}%
  \renewcommand{\tablename}{Table}%
}

\newtheorem{definition}{Definition}

\newtheorem{lemma}{Lemma}
\newtheorem{assumption}{Assumption}
\newtheorem{theorem}{Theorem}
\newtheorem{remark}{Remark}
\newenvironment{myproof}[1][\quad \it Proof: ]{{\quad \it #1: }}{\hfill $\blacksquare$\par}

\usepackage{tikz}
\usetikzlibrary{calc, graphs, intersections}
\tikzset{> = latex}
\usetikzlibrary{positioning, shapes.geometric}
\usetikzlibrary{positioning,calc}
\usetikzlibrary{backgrounds}
\usetikzlibrary{patterns}

\usepackage{pgfplots}
\usepgfplotslibrary{groupplots}
\pgfmathdeclarefunction{gauss}{3}{%
  \pgfmathparse{1/(#3*sqrt(2*pi))*exp(-((#1-#2)^2)/(2*#3^2))}%
}
\pgfmathdeclarefunction{gamma}{3}{%
  \pgfmathparse{1/(#3^#2*2)*#1^(#2-1)*2.71828^(-#1/#3)}%
}
\usepgfplotslibrary{fillbetween}
\pgfplotsset{compat=1.12} 
\def\N{50}

\usepackage{color}

\allowdisplaybreaks

\begin{document}

\title{One-Bit Model Aggregation for Differentially Private and Byzantine-Robust Personalized Federated Learning}

\author{%
	Muhang Lan, Song Xiao, and Wenyi Zhang
	\thanks{M. Lan is with School of Cyber Science and Technology, University of Science and Technology of China, Hefei, China (mhlan@mail.ustc.edu.cn); S. Xiao is with Beijing Electronic Science and Technology Institute, Beijing, China (xiaosong@mail.xidian.edu.cn); W. Zhang is with Department of Electronic Engineering and Information Science, University of Science and Technology of China, Hefei, China (wenyizha@ustc.edu.cn).}

	\thanks{This work is supported by the Natural Science Foundation
    of China (NSFC) under Grants 62231022 and 62476013.}
}




\maketitle

\pagestyle{empty}
\thispagestyle{empty}

\begin{abstract}
    As the scale of federated learning (FL) systems expands, their inherent  performance limitations like communication overhead, Byzantine vulnerability, and privacy leakage have become increasingly critical. 
    This paper considers a personalized FL framework based on model regularization, and proposes a model aggregation algorithm named PRoBit+ to concurrently overcome these limitations.
    PRoBit+ employs one-bit stochastic quantization and maximum likelihood estimation for parameter aggregation, and dynamically adjusts the step size of parameter updates, improving training stability of deep neural networks under low communication overhead and heterogeneous data distributions.
    PRoBit+'s statistical analysis is then conducted and its Byzantine robustness is proved.
    The $(\epsilon,0)$-differential privacy and a convergence upper bound of the PRoBit+ based FL are also theoretically established in heterogeneous contexts. The analysis illustrates the trade-off among transmission accuracy, security guarantees, and convergence rates, and also indicates that the performance degradation caused by transmission errors and privacy protection can be progressively eliminated at a rate of $\mathcal{O}(1/M)$ as the number of uploading clients $M$ increases. 
    Comprehensive numerical experiments are conducted to assess PRoBit+ in comparison to benchmark methods across different Byzantine attacks and varying proportions of malicious clients. 
    The experimental results demonstrate that PRoBit+ exhibits improved Byzantine robustness over existing bit-based transmission schemes, minimal performance degradation related to privacy protection, and nearly identical performance to full-precision FedAvg in a secure environment.
\end{abstract}

\begin{IEEEkeywords}
    Byzantine robustness, differential privacy, federated learning, model aggregation, maximum likelihood estimation, one-bit transmission
\end{IEEEkeywords}

\section{Introduction}
Federated learning (FL) has emerged as a promising distributed machine learning paradigm, enabling multiple clients to collaboratively train a global model under the coordination of a parameter server \cite{McMahan2017}. Since the clients keep their local datasets private and only share their model updates with the parameter server, FL provides a solution to scenarios where privacy concerns or data sovereignty issues prevent direct data sharing, such as in healthcare, finance, and mobile applications \cite{Zhang2022a,Wang2023}.
However, when FL systems scale with an increasing number of participating clients -- say, millions of mobile devices -- several critical challenges become pronounced \cite{Zhang2021}.

A widely recognized challenge is \textbf{data heterogeneity}.
Since the local data distributions of clients may vary greatly, data heterogeneity becomes severe when an FL system scales, causing the aggregated updates to converge more slowly towards the optimal global model performance \cite{Ye2023,Zhang2021a}.
As one of the representative FL methods to handle data heterogeneity, personalized FL mitigates model biases by training personalized models, leading to improved performance for individual users \cite{Tan2023,Zhang2025,Liu2024}.
Compared to other methods, personalized FL offers improved fairness and increased client engagement, and thus serves as a preferable framework for large-scale FL \cite{Lin2022}.

Another multifaceted  challenge in large-scale FL systems is model transfer.
Therein, the first prominent issue is the considerable \textbf{communication overhead}. The participation of potentially thousands or millions of clients in each training round can lead to significant bottleneck in model transmission between clients and the central server \cite{Chen2024}. Besides, in typical FL scenarios utilizing deep learning models, the size of model parameters can reach hundreds of megabytes or even gigabytes \cite{puppala2024flash}. The repeated transfer of such large models results in noteworthy latency, heightened bandwidth consumption, and potential degradation of the  overall FL efficiency.

Beyond data heterogeneity and communication overhead, security concerns also intensify with an increasing number of clients, because the probability of encountering malicious or curious clients rises \cite{Zhang2024}. Malicious clients may conduct \textbf{Byzantine attacks} \cite{Li2024a} by intentionally uploading corrupted model updates to disrupt the training process, posing a serious threat to the robustness of an FL system.
On the other hand, curious clients may attempt to infer private information from the shared model updates, leading to \textbf{privacy leakage} \cite{Yin2021}.

Given these model transfer issues, addressing them in isolation is no longer adequate; hence, it is crucial to find a training mechanism that can concurrently tackle the aforementioned problems.
SignSGD \cite{Bernstein2018} has emerged as a seminal algorithm employing signal processing (SP) principles to simultaneously enhance communication efficiency and provide Byzantine robustness.
Specifically, signSGD only transmits the signs of parameters during the upload process, resulting in a 32-fold reduction in communication costs compared to full-precision parameter uploads. Moreover, because malicious clients cannot manipulate the parameter magnitudes in signSGD, it is also Byzantine fault tolerant \cite{Bernstein2019}. 
Building upon signSGD, several SP approaches have been employed on sign-based compression to mitigate convergence failure in heterogeneous FL \cite{Chen2020}, including magnitude-driven stochastic quantization \cite{Jin2024b}, noisy perturbation \cite{Tang2024}, and sparsification \cite{Park2023}.
These methods, nevertheless, do not offer significant improvements in parameter aggregation from a SP perspective, and privacy protection has also not been sufficiently addressed.
Therefore, bit transmission aggregation mechanism is still a timely and relevant topic worthy of research in FL studies.

This paper aims to jointly address the inevitable issues of communication overhead, Byzantine vulnerability, and privacy leakage in large-scale FL systems.
To better handle device heterogeneity among clients, we employ a personalized FL framework that allows local model parameters to deviate from the global model \cite{Li2020b}.
Within this FL setup, our key contributions are as follows:
\begin{itemize}
    \item  We propose PRoBit+ (\textbf{P}rivacy-preserving and \textbf{Ro}bust \textbf{Bit} \textbf{Aggregation}), an approach based on maximum likelihood (ML) estimation whose aggregation error can be diminished as the FL system scales. It employs a stochastic one-bit compressor to decrease the communication overhead in transmitting model differences, and dynamically modifies the step size of aggregated model updates to facilitate improved training stability and faster FL convergence. Byzantine robustness is achieved since the bit-based uploading is immune to malicious parameter magnitudes, and $(\epsilon, 0)$-differential privacy is also proven by regulating the level of randomness in the binarization process.
    \item We integrate PRoBit+ into the personalized FL framework and analyze the resulting convergence behavior, deriving an upper bound for the convergence rate.
    This analysis describes how data heterogeneity, Byzantine robustness, and privacy protection interact to influence the training rate.
    It also reveals that in PRoBit+ based FL with a fixed Byzantine proportion, the performance degradation stemming from transmission errors and imposed privacy constraints can be asymptotically mitigated at a rate of $\mathcal{O}(1/M)$, where $M$ denotes the number of uploading clients.
    \item We conduct experiments involving training convolutional neural networks (CNNs) on FMNIST dataset and ResNet-18 on CIFAR-10 dataset with heterogeneous data distributions.
    The results demonstrate that the one-bit PRoBit+ aggregation can achieve recognition accuracy comparable to FedAvg with full-precision transmission in a Byzantine-free FL system.
    Furthermore,  when training deep NNs against various Byzantine attacks, the PRoBit+ consistently outperforms existing bit aggregation methods in terms of  FL performance, while also preserving client privacy without notably degrading training efficiency.

\end{itemize}
This paper is organized as follows: Section \ref{related_work} summarizes related work. Section \ref{system_model} formulates the system model and  Section \ref{aggregation_sec} describes the proposed PRoBit+ aggregation method. Sections \ref{security_sec} derives PRoBit+'s properties and presents an FL convergence analysis. Section \ref{numerical_sec} presents numerical experiment results. Finally, Section \ref{conclusion_sec}I concludes the work.

\section{Related Work}\label{related_work}

\textbf{Quantization in FL}

In FL systems, quantization, acting as a data compression technique, is employed to mitigate bandwidth constraints during model transmission.
In \cite{Konecny2016}, quantization with random rotations and subsampling is integrated to improve the communication process.
In \cite{Alistarh2017}, a compression scheme termed QSGD quantizes each component to a discrete set of values via randomized rounding, thereby preserving the statistical properties of the original gradients. SignSGD adopts a more aggressive approach by quantizing gradients to their sign only, significantly reducing the communication overhead \cite{Bernstein2018}.
A vector quantization scheme called UVeQFed is proposed to enhance efficiency, assuming a shared source of common randomness between clients and server \cite{Shlezinger2021}. 

To improve upon universal schemes, adaptive methods have been proposed for fine-grained quantization, adjusting quantization parameters based on evolving statistics of gradients \cite{Faghri2020}, local data distribution \cite{Mao2022}, and varying parameter significance across clients \cite{Li2024}.
From the perspective of network scheduling, quantization-level scheduling methods based on reinforcement learning have also been proposed to jointly optimize the learning performance and latency \cite{Zhang2024b,Zhang2024c}.

Clearly, the research focus in FL quantization is shifting from fixed-bit stochastic quantization to dynamically adjusted mixed-precision quantization. 
PRoBit+ integrates these two attributes: To minimize communication overhead in large-scale FL systems, PRoBit+ still employs global one-bit quantization; meanwhile, PRoBit+ dynamically modifies the level of quantization randomness during the training process, leading to faster convergence rates, as corroborated in numerical experiments.

\textbf{Differential Privacy}

Differential privacy (DP) provides a rigorous framework for quantifying privacy leakage by precisely controlling information disclosure. DP guarantees that an algorithm's output remains consistent even when the input data varies by a single record \cite{dwork2014algorithmic}. This is typically achieved by introducing randomness into the data processing pipeline, often through the addition of noise during algorithmic computations.

In the realm of FL, DP plays a critical role in protecting against the deductive disclosure of individual data from model updates \cite{Mothukuri2021}.
Inspired by standard additive DP mechanisms, calibrated Gaussian noise is added to the average of clipped local updates in \cite{Geyer2017}.
This approach has also been extended to quantized updates, employing different noise distributions such as Binomial \cite{Agarwal2018}, Skellam \cite{Agarwal2021}, and discrete Gaussian \cite{Kairouz2021b} to achieve simultaneous quantization and privacy.
However, these noise-addition-based methods lead to biased estimation owing to the demand of modular clipping. 

Subsequent studies then leverage compression techniques with inherent randomness to attain privacy without incorporating additive noise.
Such randomized compression schemes include canonical sketches  \cite{Li2019a}, random lattice \cite{Lang2023}, randomized sub-sampling and rounding \cite{Youn2023} and randomized response based quantization schemes \cite{Gandikota2022}.

These studies reveal the significance of randomness in quantization-based DP mechanisms. Through dynamic adjustment of stochasticity in bit quantization, PRoBit+ achieves both DP and fine-grained control over FL performance-privacy tradeoff.

\textbf{Byzantine-robustness in Heterogeneous FL}

Early studies on Byzantine robustness primarily concentrate on homogeneous data settings, with key contributions from clustering-based algorithms \cite{Blanchard2017, ElMhamdi2018,Xia2019},  median-based algorithms \cite{Yin2018,Pillutla2022}, and coding-based algorithms \cite{Chen2018,Rajput2019}. 
In \cite{Bernstein2019},  signSGD is shown to maintain robust convergence even when up to $50\%$ of nodes are under adversarial conditions. While these robust aggregation methods leverage statistical consensus among benign clients to identify outliers, they struggle with heterogeneous data, which obscures the distinction between data-induced and attack-induced anomalies.

This limitation has motivated the development of Byzantine-tolerant algorithms specifically designed for heterogeneous data distributions.
These include bucketing aggregation \cite{Karimireddy2020a} and gradient splitting \cite{Liu2023} for mitigating data heterogeneity, as well as  multi-armed bandit based client selection \cite{Wan2022} and iterative clipping \cite{Karimireddy2021} for identifying honest clients via their training history. Additionally, $l_p$ norm regularization has also been explored as an optimization-based defense against Byzantine attacks \cite{Li2019}.

However, designing Byzantine-robust methods for heterogeneous data distributions under one-bit model transmission remains a relatively unexplored area. Prevailing methods typically rely on the inherent robustness of signSGD, employing majority vote \cite{Ma2021,Safaryan2021} or sign accumulation \cite{zhu2022bridging,Jin2024b} for aggregation, suffering from training instability due to their reliance on manually tuned step sizes.
In our work, PRoBit+ mitigates this issue by employing an adaptive step size for server model updates, thus enabling stable training of deep NNs even with heterogeneous data and one-bit communication.

\section{System Model}\label{system_model}
In this section, we start with a standard FL model and modify the FedAvg algorithm to establish the training framework used in this paper.

\subsection{FL Preliminaries}
We consider a standard FL framework \cite{McMahan2017}, within which a parameter server and multiple clients collaboratively solve a supervised learning problem. Client $ m $ owns a private dataset $ \mathcal{D}_m $.
Without loss of generality, in this paper we assume that the sizes of the datasets from different clients are equal, denoted as
$ D$.  Loss function $ l(\bm w ; \bm \xi) $ measures how well a machine learning model with parameter $ \bm w\in \mathbb{R}^{d} $ fits a particular data sample $ \bm \xi $, and the loss function of client $ m $ is defined as
\begin{equation}
	f_m(\bm w) \triangleq \frac{1}{D} \sum_{\bm \xi \in \mathcal{D}_m} l(\bm w; \bm \xi). \label{local_loss}
\end{equation}
Assume that there are $M$ clients participating in FL, where $ R $ of them are regular clients denoted by set  $\mathcal{R}$ and the rest are Byzantine clients denoted by set $\mathcal{B}$, with their identities unknown to the server.
FL aims at finding optimal models to minimize the global loss function on $R$ regular clients, formulated as
\begin{align}
    \min_{\bm W} ~ & F(\bm w)= \frac{1}{R}\sum_{m=1}^{R}   f_m(\bm w^m), \label{flmodel} \\
    \mathrm{s.t.} ~ & \bm w^m = \bm w, ~ \forall m \in \mathcal{R}, \label{constraints}
\end{align}
where $\bm W:=[\cdots;\bm w^{m};\cdots;\bm w]\in\mathbb{R}^{d\times (R+1)}$ stacks $R$ local models $\{\bm w^m\}_{m \in \mathcal{R}}$ on the regular clients and one global model $\bm w$ on the parameter server.
Throughout this paper, we use superscript $m$ to denote the client index.

The FedAvg algorithm \cite{McMahan2017} has been proposed to solve the FL problem in the form of \eqref{flmodel}, through iteratively executing the following steps at each round.
In round $t$, the parameter server broadcasts its current global model $\bm w_t$ to all clients.
Upon receiving the global model, client $m$ uses it as the initial point to train its local model, aiming to approximately minimize the local loss function \eqref{local_loss}, and resulting in the local model $\bm w^m_{t+1}$.
Next, the clients synchronously upload their model differences $\bm \delta^{m}_t \triangleq \bm w^m_{t+1}-\bm w_t$ to the parameter server.
The parameter server averages the model differences via $\bm \theta_t = \frac{1}{M}\sum_{m=1}^{M} \bm \delta^{m}_t $, and updates the global model as $ \bm w_{t+1} = \bm w_{t}+\bm \theta_t$.
Then, the algorithm proceeds to the next round and completes the training process after $T$ rounds. 
By indiscriminately incorporating all client updates -- including those deliberately engineered by Byzantine clients -- into its averaging process without mitigation, FedAvg suffers a pronounced Byzantine vulnerability.
We address this issue by leveraging the robustness of bit aggregation against amplitude manipulation, as detailed in Section \ref{aggregation_sec}.

\subsection{Model Regularization}
FedAvg, while exploiting the aggregation properties of deep NNs to mitigate model drift and accelerate FL training, suffers from poor convergence on highly heterogeneous data.
Specifically, starting each local training round with an identical global model restricts model diversity, preventing FL clients from effectively learning local data and, as a result, reducing overall performance.
Therefore, we exploit a personalized FL technique named model regularization to address these two challenges \cite{Li2020b,Shi2023}.
Specifically, we relax the constrained optimization problem \eqref{flmodel} \eqref{constraints} to obtain a new optimization problem:
\begin{align*}
    & \min_{\bm W} ~ \frac{1}{R}\sum_{m=1}^{R} f_m(\bm w^m)+\frac{\lambda}{2}\left\|\bm w^m-\bm w\right\|^2,
\end{align*}
where the squared $l_2$-norm penalty term parameterized by $\lambda > 0$ forces the local models $\{\bm w^m\}_{m \in \mathcal{R}}$ to be close to $\bm w$, while still allowing them to differ to enable personalization. The optimization objective of local training at each client is modified accordingly into
\begin{align}
    \min_{\bm w^m} h_m(\bm w^m;\bm w) = f_m(\bm w^m)+\frac{\lambda}{2}\left\|\bm w^m-\bm w\right\|^2.
    \label{eq3-model3}
\end{align}

\subsection{Secure Compressed Transmission}

In addition to the aforementioned challenges, FedAvg also exhibits deficiencies in communication efficiency and security. To address this issue, we propose a secure compressed transmission scheme named  PRoBit+, which compresses model updates into bits $\bm c_t^m = \text{Compress}(\bm \delta^{m}_t)$ and aggregates them into $\hat{\bm \theta}_t \triangleq \text{Aggregate} \left(\bm c^{1}_t,\cdots,\bm c^{M}_t\right)$ at the server.
Here, $ \text{Compress}(\cdot)$ and $ \text{Aggregate}(\cdot)$ refer to general operators, and a detailed illustration to the proposed algorithm and an aggregation analysis are presented in Sections \ref{aggregation_sec} and \ref{security_sec}, respectively.
The subscript $t$ will be omitted in the context of a single transmission round.

\section{FL Training Scheme with PRoBit+}\label{aggregation_sec}
In this section, we first introduce the two components of PRoBit+: the bit compressor of model updates and the ML-based parameter aggregation on the server side. Then, we present the overall FL training mechanism with PRoBit+.

\subsection{Bit Compressor}
To reduce FL communication overhead, parameter bit quantization is commonly employed. 
However, current threshold-based one-bit quantization discards magnitude information, hindering FL performance under heterogeneous data distributions \cite{Jin2024}.
Therefore, we employ a stochastic quantizer \cite{Jin2024b} to preserve the magnitude information of the parameters. 
Instead of directly transmitting the sign bits of the parameters, our two-level stochastic quantizer assigns probabilities to different quantization outcomes based on parameter magnitudes, with larger values more likely to be quantized to $ 1 $ and smaller values more likely to be quantized to $ -1 $.
This stochastic quantizer can be viewed as a special case of QSGD operating within the quantization range of $[-1, 1]$ \cite{Alistarh2017}, enabling the transmission of magnitude information without incurring additional communication overhead.
Specifically, for any given model update $\bm \delta^{m}$, the stochastic quantization compressor outputs $\bm c^{m}=\text{Compress}(\bm \delta^{m},\bm b)$, where the $i$-th component is:
\begin{align}
    c_i^m=\begin{cases}1, & \text{with probability}\frac{b_i+\delta^{m}_{i}}{2b_i},
        \\-1,&\text{with probability }\frac{b_i-\delta^{m}_{i}}{2b_i},\end{cases}
    \label{stosign}
\end{align}
where $\bm b$ is a pre-designed parameter vector satisfying $b_i\geq max_m|\delta^{m}_{i}|$.

\subsection{ML-based Aggregation}

While model differences among clients are permitted in personalized FL, server-side model aggregating remains essential as it stabilizes and mitigates model drift during FL training. 
Given that the server is unaware of both the number and identities of Byzantine clients, we initially consider a Byzantine-free transmission scenario and derive an aggregation method based on ML estimation. In this scenario, FedAvg is among the simplest and most effective methods for FL aggregation, making it a suitable reference for one-bit aggregation.

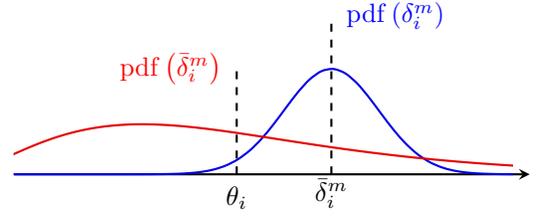
\begin{figure}
  \centering
  \begin{tikzpicture}[]
      \message{Bias & systematic error^^J}
      \def\B{8};
      \def\T{4};
      \def\V{6};
      \def\Bs{1.0};
      \def\Ts{1.0};
      \def\Vs{2.0};
      \def\xmax{\B+3.2*\Bs};
      \def\ymin{{-0.3*gauss(\B,\B,\Bs)}};
      \def\ymax{{1.1*gauss(\B,\B,\Bs)}};

      \begin{axis}[
          every axis plot post/.append style={
          mark=none,domain={-0.05*(\xmax)}:{1.08*\xmax},samples=\N,smooth},
          xmin={1.3}, xmax=\xmax,
          ymin=\ymin, ymax=\ymax,
          axis lines=middle,
          axis line style=thick,
          enlargelimits=upper,
          ticks=none,
          every axis x label/.style={at={(current axis.right of origin)},anchor=north west},
          y=200pt,
          axis y line=none
          ]

          \addplot[name path=B,thick,black!10!blue ] {0.5*gauss(x,\B,\Bs)};
          \addplot[name path=S,thick,black!10!red  ] {0.7*gamma(x,3,2)};
          \addplot[black,dashed,thick]
          coordinates {(\B,0) (\B, 0.3)}
          node[below=-2pt,pos=0] {$\bar{\delta}^{m}_i$}
          node[above=1pt,right=2pt,color = blue] {$\mathrm{pdf}\left(\delta^{m}_i\right)$};

          \addplot[black,dashed,thick]
          coordinates {(6,0) (6, 0.2)}
          node[below=1pt,pos=0] {$\theta_i$}
          node[above=0pt,left=2pt,color = red] {$\mathrm{pdf}\left(\bar{\delta}^{m}_i\right)$};

      \end{axis}
  \end{tikzpicture}
  \caption{Distribution of client updates $\delta^{m}_i$ and their means $\bar{\delta}^{m}_i$.  }
  \label{distri_delta}
\end{figure}

To distributedly estimate the mean of parameters under heterogeneous distributions, it is inevitable to pose assumptions on the correlation among different clients' data distributions. For this purpose, we make statistical assumptions concerning the parameters of different clients at the same entry: each component  $\delta^{m}_i$ follows an independent Gaussian distribution \cite{gao2019rate}, and their means $\bar{\delta}^{m}_i$ conform to a same distribution, as illustrated in Fig.  \ref{distri_delta}. This assumption simultaneously addresses the similarities and sampling biases in heterogeneous FL data.

In this case, the FedAvg expression coincides with the ML estimate of the mean parameter $\theta_{i}$, provided that $\frac{1}{M}\sum_{m=1}^{M} \bar{\delta}^{m}_i$ serves as the ML estimate of $\theta_{i}$, e.g., when $ \theta_{i}$ follows a common exponential family distribution.
To approximate the FedAvg aggregation result, we re-derive the ML estimate of $\theta_{i}$ within the context of the bit compressor \eqref{stosign}.
Initially, we have
\begin{align}

    \mathbb{P}(c^1_{i},c^2_{i},\cdots,c^M_{i})

     & = \prod_{m=1}^{M} \int_{-b_i}^{b_i} \mathbb{P}(c^m_{i}|\delta_i^m)\mathrm{d}F(\delta_i^m)   \label{int}            \\

     & =  \prod_{m=1}^{M} \int_{-b_i}^{b_i} \mathbb{P}(c^m_{i}|\delta_i^m)f(\delta_i^m)\mathrm{d}\delta_i^m, \label{eq3-9}

\end{align}
where \eqref{int} represents the Riemann–Stieltjes integral and $F(\delta_i^m)$ is the cumulative distribution function of the random variable $\delta_i^m$.
Calculating the probability that $c^m_i$ takes the value of 1, we have
\begin{align}

    \mathbb{P}(c^m_i=1)

     & = \int_{-b_i}^{b_i} \mathbb{P}(c^m_i=1|\delta_i^m)f(\delta_i^m)\mathrm{d}\delta_i^m      \notag                         \\

     & \approx \int_{-\infty}^{\infty} \frac{b_i+\delta_i^m}{2b_i} f(\delta_i^m)\mathrm{d}\delta_i^m \label{eq3-5}                               \\

     & = \frac{1}{2}+\frac{1}{2b_i}\mathbb{E}\left[\delta_i^m\right]  \notag                \\

     & = \frac{1}{2}+\frac{1}{2b_i}\mathbb{E}\left[\bar{\delta}^{m}_i\right]         \label{expectation}                              \\

     & = \frac{1}{2}+\frac{1}{2b_i}\theta_{i}, \label{eq3-6}

\end{align}
where \eqref{eq3-5} is approximated under the assumption of a Gaussian distribution for $\delta_i^m$, which is not long-tailed; \eqref{expectation} is derived from the total probability formula.
Similarly, we can obtain
\begin{align}

    \mathbb{P}(c^m_i=-1)=  \frac{1}{2}-\frac{1}{2b_i}\theta_{i}. \label{eq3-7}

\end{align}
Substituting \eqref{eq3-6} and \eqref{eq3-7} back into \eqref{eq3-9} yields
\begin{align}

    &\mathbb{P}(c^1_{i},c^2_{i},\cdots,c^M_{i};\theta_i)  \notag \\ 
    = {} &  \left(\frac{1}{2}+\frac{1}{2b_i}\theta_{i}\right)^{N_i}\left(\frac{1}{2}-\frac{1}{2b_i}\theta_{i}\right)^{M-N_i}, \label{probability_func}

\end{align}
where $N_i=\sum_{m=1}^{M}\mathbb{I}\left\{c^m_{i}=1\right\}$ represents the number of $1$ in the set $\{c^m_{i}\}$.

By the method of ML estimation, we have $\hat{\theta}_{i} = \arg \max_{\theta_i} \mathbb{P}(c^1_{i},c^2_{i},\cdots,c^M_{i};\theta_i)$, and the estimated value is obtained as 
\begin{align}
    \hat{\theta}_{i} = \frac{2N_i-M}{M}b_i.
\label{ml-aggre}
\end{align}
We adopt this estimation result as the aggregated parameter on the server side and extend it to the scenario with malicious users. Then the  global parameter is updated as $ \bm w_{t+1} = \bm w_{t}+\hat{\bm \theta}_t$.

\begin{remark}
    Current prevalent one-bit aggregation methods can be classified into two main categories. One category is majority vote \cite{Bernstein2019,Safaryan2021,Ma2021}, which quantifies the occurrences of $1$ and $-1$ in the uploaded bits, selecting the value with the higher frequency as the aggregation result. 
    The other category further considers the relative quantity relationship between the two bit values, directly summing the received bits of $1$ and $-1$ to generate the aggregation result \cite{Jin2024b,zhu2022bridging}. 
    However, to mitigate the negative impact of bit aggregation on FL numerical stability, both methods require additional manual adjustment of the update step size, thereby constraining the convergence rate of the learning process. Our proposed PRoBit+ bit transmission algorithm, utilizing ML estimation, effectively addresses this issue.
\end{remark}

\subsection{FL Training Scheme}

The aforementioned stochastic bit compressor, combined with the ML-based parameter estimation, constitutes the PRoBit+ algorithm. By further integrating model regularization into local training, the overall FL training mechanism is established, as seen in Algorithm \ref{robust_algo}.

Unlike FedAvg, which mitigates model drift by initiating local training directly from the received global model $\bm w_t$, our proposed method employs model regularization to suppress model drift, enabling clients to maintain their local parameters and initiate local training from $\bm w^m_t$, thereby enhancing the personalization of client parameters. This approach ultimately leads to a final aggregated model $\bm w_T$ that exhibits improved performance on heterogeneous data.  
Furthermore, single-bit transmission enhances communication efficiency; the randomized mapping \eqref{stosign} preserves client privacy; the magnitude-independent transmission scheme fortifies Byzantine robustness. Rigorous analyses are presented in the following section.

\begin{algorithm}[t]
    \caption{Proposed FL training scheme with PRoBit+}\label{robust_algo}
    \KwIn{Quantization parameter $\bm b$, initial server parameter $\bm w$;}
    \KwOut{Server parameter $\bm w_{T}$;}
    \For{$t = 0, 1, \cdots,T$}{
        \textbf{Server:}

        Broadcast $\bm w_{t}$ to all clients\;
        Receive compressed updates $\bm c^m_t$ from clients\;
        Compute aggregated model update $\hat{\bm \theta}_t$ via \eqref{ml-aggre}\;
        Update global parameter $\bm w_{t+1} = \bm w_{t}+\hat{\bm \theta}_t$\;
        \textbf{Client:}

        Receive $\bm w_{t}$ from server\;
        Locally train $\bm w^m$ to solve \eqref{eq3-model3} and obtain model update $\bm \delta^{m}_t = \bm w^m_{t+1}-\bm w_t$\;
        Apply stochastic bit compressor \eqref{stosign} to obtain $\bm c^m_t$\;
        Send $\bm c^m_t$ to server\;
    }
  \end{algorithm}

\section{Theoretical Analysis}\label{security_sec}

In this section, we first analyze the statistical properties of the PRoBit+ aggregation and prove its Byzantine robustness. Then, focused on PRoBit+ based FL, we demonstrate its DP guarantees and conduct convergence analysis. Finally, we explore how the PRoBit+ parameter $\bm b$ balances transmission accuracy, Byzantine robustness, privacy, and convergence rate.

\subsection{PRoBit+ Aggregation}\label{aggre_property}
\subsubsection{Statistical Properties}
We analyze the statistical properties of the PRoBit+ aggregation in the absence of Byzantine attacks.
By first proving that $\hat{\bm{\theta}}$ is a sufficient statistic and an unbiased estimate of $\bm{\theta}$, we demonstrate the validity of PRoBit+. Additionally, we analyze its estimation error, showing that increasing the number of clients can reduce the transmission error at a rate of $\mathcal{O}\left(1/M\right)$.
Therefore, PRoBit+ could serve as an incentive mechanism that encourages numerous clients to share their model updates, enhancing overall FL performance.
Above properties are detailed in Theorem \ref{statistic_theorem}, whose proof is provided in Appendix \ref{statistic_theorem-proof} in the supplementary material.

\begin{theorem} 
    Under Byzantine-free conditions, the PRoBit+ aggregation result $\hat{\bm{\theta}}$ exhibits following properties:
    \begin{enumerate}
        \item $\hat{\bm{\theta}}$ is a sufficient statistic for $\bm{\theta}$.
        \item $\hat{\bm{\theta}}$ is an unbiased estimate of $\bm{\theta}$, i.e., $\mathbb{E}[\hat{\bm{\theta}}]=\bm{\theta}$.
        \item The estimation error is
        \begin{align}
            \mathbb{E}\left[\left\|\bm{\theta}-\hat{\bm{\theta}}\right\|^2\right]=\frac{\sum_{i=1}^{d}\left(b_i^2-\theta_i^2\right)}{M}.
            \notag 
        \end{align}
    \end{enumerate}
    \label{statistic_theorem}
\end{theorem}

\subsubsection{Byzantine Robustness}

Byzantine clients are assumed to be omniscient, colluding, and capable of sending arbitrary malicious updates to the server to bias the learning process.
By restricting each client to uploading only one bit, PRoBit+ effectively alleviates the impact of malicious messages.  
We prove in Theorem \ref{theorem-1} that the PRoBit+ result is only affected by the proportion of Byzantine clients, rather than the magnitudes of malicious updates.
Consequently, PRoBit+ exhibits robustness against arbitrary attacks, especially for those involving unbounded or zero-valued updates. This is in sharp comparison with FedAvg, which is vulnerable to even a single Byzantine client.

\begin{theorem}\label{theorem-1}
    Denote the proportion of Byzantine clients as $\beta$, then the deviation induced by any Byzantine attack on the PRoBit+ aggregation result is upper bounded as
\begin{align}
    \left\|\mathbb{E}\left[\bm{\theta}\right]_{R}-\mathbb{E}\left[\bm{\theta}\right]_{B}\right\| \le 2\beta\left\|\bm{b}\right\|,
    \notag 
\end{align}
where $\mathbb{E}\left[\bm{\theta}\right]_{R}$ and $\mathbb{E}\left[\bm{\theta}\right]_{B}$ denote the aggregated value without and with Byzantine attacks, respectively.
\end{theorem}
\begin{myproof}[Proof of Theorem \ref{theorem-1}]
    We denote the expected value of $N_i$ without Byzantine attacks as $\mathbb{E}\left[N_i\right]_{R}$, and decompose it as follows:
    \begin{align}
        \mathbb{E}\left[N_i\right]_{R} = \sum_{m\in \mathcal{R}}\mathbb{P}\left(c_i^m=1\right)_{R} +\sum_{m\in\mathcal{B}}\mathbb{P}\left(c_i^m=1\right)_{R}, \notag 
    \end{align}
    where $\mathbb{P}\left(c_i^m=1\right)_{R}$ represents the honest mapping probability from $\delta_i^m$.
    Similarly, in the presence of Byzantine clients, this statistic is modified as
    \begin{align*}
        \mathbb{E}\left[N_i\right]_{B} = \sum_{m\in \mathcal{R}}\mathbb{P}\left(c_i^m=1\right)_{R} +\sum_{m\in\mathcal{B}}\mathbb{P}\left(c_i^m=1\right)_{B}. 
    \end{align*}
    Since the probability value does not exceed $1$, we have an upper bound
    \begin{align*}
        \mathbb{E}\left[N_i\right]_{B} &\le  \sum_{m\in \mathcal{R}}\mathbb{P}\left(c_i^m=1\right)_{R} +\beta M \\
        &\le  \sum_{m\in \mathcal{R}\bigcup\mathcal{B}}\mathbb{P}\left(c_i^m=1\right)_{R} +\beta M \\
        &= \mathbb{E}\left[N_i\right]_{R} +\beta M.
    \end{align*}
    Similarly, we can obtain a lower bound as
    \begin{align*}
        \mathbb{E}\left[N_i\right]_{B} &\ge  \sum_{m\in \mathcal{R}}\mathbb{P}\left(c_i^m=1\right)_{R} \\
        &\ge  \sum_{m\in \mathcal{R}\bigcup\mathcal{B}}\mathbb{P}\left(c_i^m=1\right)_{R} -\beta M \\
        &= \mathbb{E}\left[N_i\right]_{R} -\beta M.
    \end{align*}
    Therefore, the aggregation bias caused by Byzantine attacks satisfies
    \begin{align*}
        \left|\mathbb{E}\left[\theta_i\right]_{R}-\mathbb{E}\left[\theta_i\right]_{B}\right| \le 2\beta b_i.
    \end{align*}
    By extrapolating this result to a vector notation, we can obtain
    \begin{align*}
        \left\|\mathbb{E}\left[\bm{\theta}\right]_{R}-\mathbb{E}\left[\bm{\theta}\right]_{B}\right\| &= \sqrt{\sum_{i=1}^{d}\left(\mathbb{E}\left[\theta_i\right]_{R}-\mathbb{E}\left[\theta_i\right]_{B}\right)^2} \\
        &\le \sqrt{4\beta^2\sum_{i=1}^{d}b_i^2} \\
        &= 2\beta\left\|\bm{b}\right\|. 
    \end{align*}
\end{myproof}

\subsection{PRoBit+ based FL}\label{dpanalysis}
\subsubsection{Differential Privacy Guarantee}

We quantify client privacy in terms of DP, whose definition is as follows \cite{dwork2014algorithmic}:
\begin{definition}
    A randomized algorithm $\mathcal{M}$ is said to provide $(\epsilon,\delta)$-DP if for all adjacent inputs $\mathcal{I}_a,\mathcal{I}_b$ ($\|{\mathcal{I}}_a-\mathcal{I}_b\|_1\leq1$) and all possible output sets $\mathcal{O}$, the following condition holds:
    \begin{align*}
        \mathbb{P}\left(\mathcal{M}(\mathcal{I}_a)\in\mathcal{O}\right)\leq e^\epsilon\mathbb{P}\left(\mathcal{M}(\mathcal{I}_b)\in\mathcal{O}\right)+\delta,
    \end{align*}
    where the privacy loss $\epsilon$ controls the trade-off between privacy and algorithm utility, and $\delta$ is the probability that $\epsilon$-DP may fail.
\end{definition}

We consider an honest-but-curious FL server, attempting to infer information about clients' private data by analyzing the received updates. 
Thus, a local randomizer should be employed to perturb the local message sent from each client. 
This model of DP, known as local DP, ensures that even if the master server or an adversary accesses the clients’ shared parameters, they still cannot extract significant information about this client’s dataset.

To mitigate this privacy risk, we propose employing a local randomizer to perturb client-side messages. This approach, leveraging local DP, guarantees that even if the server or a malicious third party gains access to the shared parameters, sensitive information about individual client datasets remains protected.

Next, we prove that the PRoBit+ transmission scheme satisfies the given $(\epsilon,0)$-DP requirements.
Note that although we analyze the per-round-of-communication privacy budget to ensure privacy, it is also feasible to exploit advanced composition theorems or the analytical moments accountant method to obtain the multi-round privacy loss \cite{abadi2016deep,dwork2014algorithmic}.
We denote the outputs of the two adjacent datasets as $\bm{\delta}^m$ and ${\bm{\delta}^m}^\prime$, respectively, and define their difference as $\bm{v}^m \triangleq {\bm{\delta}^m}^\prime-\bm{\delta}^m$. The vector $\bm{v}^m$ satisfies $\left\|\bm{v}^m\right\|_1 \le \Delta_1$, where $\Delta_1$ is the $l_1$-sensitivity of $\bm{\delta}^m$, quantifying the maximum output change induced by altering a single element in the dataset.
As shown in \eqref{stosign}, the quantizer's randomness is governed by the hyperparameter $\bm b$. 
Increasing the value moves the probability mass of the quantized parameter's two-point distribution closer to $1/2$, thus enhancing DP.
Building on this insight, we prove that the PRoBit+ based FL satisfies $(\epsilon,0)$-DP, as detailed in Theorem~\ref{dptheorem}.
\begin{theorem}
    The PRoBit+ based FL satisfies $(\epsilon,0)$-DP when setting $ b_i  \ge \max_m|\delta^{m}_{i}|+\left(1+\frac{1}{\epsilon}\right)\Delta_1 $.
    \label{dptheorem}
\end{theorem}
\begin{myproof}[Proof of Theorem \ref{dptheorem}]
    According to the definition of privacy loss \cite{dwork2014algorithmic}, we have
    \begin{align*}
        PL & = \ln \frac{\mathbb{P}\left(\bm c^m|\bm \delta^m+\bm v^m\right)}{\mathbb{P}\left(\bm c^m|\bm \delta^m\right)}         \\
           & = \sum_{i=1}^{d} \ln \frac{\mathbb{P}\left(c^m_i|\delta^m_i+ v^m_i\right)}{\mathbb{P}\left( c^m_i| \delta^m_i\right)},
    \end{align*}
    where $\bm c^m$ represents the values after stochastic quantization.
    We further analyze the privacy loss for the $i$-th dimension. Consider the case when $c^m_i=1$, we have
    \begin{align*}
        PL _i & =  \ln \frac{\mathbb{P}\left(c^m_i=1|\delta^m_i+ v^m_i\right)}{\mathbb{P}\left(c^m_i=1|\delta^m_i\right)}        \\
              & = \ln \frac{\left(b_i+\delta^m_i+ v^m_i\right)/\left(2b_i\right)}{\left(b_i+\delta^m_i\right)/\left(2b_i\right)} \\
              & = \ln \left(1+ \frac{v^m_i}{b_i+\delta^m_i}\right)                                                                \\
              & \le  \frac{v^m_i}{b_i+\delta^m_i}. 
    \end{align*}
    From the given condition on $b_i$, it holds
    \begin{align*}
        b_i \ge \max_m|\delta^{m}_{i}|+\left(1+\frac{1}{\epsilon}\right)\Delta_1   \ge -\delta^{m}_{i} + \frac{\Delta_1}{\epsilon}.
    \end{align*}
    By simple manipulation of the equation, we can obtain
    \begin{align*}
        \frac{1}{b_i+\delta^m_i} \le  \frac{\epsilon}{\Delta_1}.
    \end{align*}
    Summing $PL_i$ over all dimensions, we get
    \begin{align*}
        \sum_{i=1}^{d} PL_i
          \le \sum_{i=1}^{d}\frac{|v^m_i|}{b_i+\delta^m_i}   
          \le \frac{\epsilon}{\Delta_1}\sum_{i=1}^{d}|v^m_i| 
          \le \epsilon, 
    \end{align*}
    where the last inequality follows from the $l_1$-sensitivity.

    Similarly, when $c^m_i=-1$, the privacy loss is
    \begin{align*}
        PL _i & =  \ln \frac{\mathbb{P}\left(c^m_i=-1|\delta^m_i\right)}{\mathbb{P}\left(c^m_i=-1|\delta^m_i+ v^m_i\right)}      \\
              & = \ln \frac{\left(b_i-\delta^m_i\right)/\left(2b_i\right)}{\left(b_i-\delta^m_i- v^m_i\right)/\left(2b_i\right)} \\
              & = \ln \left(1+ \frac{v^m_i}{b_i-\delta^m_i- v^m_i}\right)                                                        \\
              & \le  \frac{v^m_i}{b_i-\delta^m_i- v^m_i}. 
    \end{align*}
    From the given condition on $b_i$, we have
    \begin{align*}
        b_i \ge \max_m|\delta^{m}_{i}|+\left(1+\frac{1}{\epsilon}\right)\Delta_1  \ge \delta^{m}_{i} +v^m_i+ \frac{\Delta_1}{\epsilon},
    \end{align*}
    which holds 
    \begin{align*}
        \frac{1}{b_i-\delta^m_i- v^m_i} \le  \frac{\epsilon}{\Delta_1}.
    \end{align*}
    Summing $PL_i$ over all dimensions, we get
    \begin{align*}
        \sum_{i=1}^{d} PL_i
          \le \sum_{i=1}^{d}\frac{|v^m_i|}{b_i-\delta^m_i- v^m_i} 
          \le \frac{\epsilon}{\Delta_1}\sum_{i=1}^{d}|v^m_i|      
          \le \epsilon.
    \end{align*}
    Combining both cases for $c^m_i=1$ and $c^m_i=-1$, we have $PL=\sum_{i=1}^{d} PL_i \le \epsilon$, thus the stochastic quantization mechanism satisfies $(\epsilon,0)$-DP.
\end{myproof}

\subsubsection{Convergence Analysis}
\label{robust_convergence}

This part expands on the previous aggregation analysis to conduct a PRoBit+ based FL convergence analysis. 
We first make assumptions regarding the data heterogeneity and analytical properties of the FL system, as described in Assumptions \ref{assup_dissimilarity} and \ref{assup_smoothness}. These assumptions are commonly utilized in the FL literature \cite{Li2020b} and enable us to conduct the convergence analysis effectively.

\begin{assumption}
    For all non-stationary points $\bm w_t  \in \{\bm w\mid\|\nabla F(\bm w)\|>\alpha\}$, the local functions $f_m$ are $B$-dissimilar, i.e., $B(\bm w_t)\le B$, where $B(\bm w)$ is defined as $B(\bm w)\triangleq\sqrt{\frac{\mathbb{E}_m[\|\nabla f_m(\bm w)\|^2]}{\|\nabla F(\bm w)\|^2}}$ when $\|\nabla F(\bm w)\|\neq 0$.
    \label{assup_dissimilarity}
\end{assumption}

\begin{assumption}
    The functions $f_m$ are non-convex, $L$-Lipschitz smooth, and $L_0$-Lipschitz continuous, and there exists $L_{-}>0$ such that $\nabla^{2}f_{m}\succeq-L_{-}\bm I$, where $\bar{\lambda}=\lambda-L_{-}>0$.
    \label{assup_smoothness}
\end{assumption}

Considering the heterogeneity in computational resources among FL clients, adopting a common number of local training rounds may pose a challenge in achieving timely model synchronization.
Therefore, instead of explicitly specifying the number of training samples or epochs, we use $\gamma$-inexactness to flexibly characterize the training process of local clients \cite{Li2020b}, as formalized in Definition \ref{inexact}. 
We also define a secure margin in Definition \ref{margin} to ensure that the quantization hyperparameter $\bm b$ satisfies $ b_i  \ge \max_m|\delta^{m}_{i}|$.

\begin{definition}[$\gamma$-Inexact Solution]
    For the function $h(\bm w;\bm w_0) = f(\bm w)+\frac{\lambda}{2}\left\|\bm w-\bm w_0\right\|^2$ and $\gamma \in [0,1]$, when $\|\nabla h(\bm w^*;\bm w_0)\| \leq \gamma\|\nabla h(\bm w_0;\bm w_0)\|$ holds, we define $\bm w^*$ as a $\gamma$-inexact solution of $\min_{\bm w} h(\bm w;\bm w_0)$, where $\nabla h(\bm w;\bm w_0)=\nabla f(\bm w)+\lambda(\bm w-\bm w_0)$. Note that a smaller $\gamma$ corresponds to higher accuracy.
    \label{inexact}
\end{definition}

\begin{definition}
    We define the security margin $\bm \zeta $ such that $ b_{t,i}=\theta_{t,i} +\zeta_i  \ge \max_m|\delta^{m}_{t,i}|, \forall t \in \left\{1,2,\cdots,T\right\}$.
    \label{margin}
\end{definition}

Next, we proceed with the FL convergence analysis as detailed in Theorem \ref{theorem-3} below.
\begin{theorem}\label{theorem-3}
    By selecting appropriate $\gamma$, $\lambda$, and $M$ to ensure the constant $\rho>0$, the convergence rate to the point $\{\bm w\mid\mathbb{E}\big[\|\nabla f(\bm w)\|\big]{\leq}\alpha\}$ satisfies
    \begin{align*}
        & \frac{1}{T}\sum_{t=1}^{T}\left\|\nabla F(\bm w_{t})\right\|^{2} \le \frac{1}{\rho}\left(\frac{F(\bm w_0)-F^\star}{T} \notag \right.\\
        & + \left. 2L_0\left(\sqrt{2}\beta\left\|\bm \zeta\right\|+\sqrt{\frac{\Delta_1}{M}\left(1+\frac{1}{\epsilon}\right)d\left\|\bm\zeta\right\|}\right)\right),
    \end{align*}
    where $\rho = \frac{1}{\lambda}-\mathcal{O}\left(B^2\right)$.
\end{theorem}
\begin{myproof}[Proof outline of Theorem \ref{theorem-3}]

    We first derive a convergence rate upper bound  in the context of lossless and Byzantine-free transmission, as stated in Lemma \ref{lemma-5}. Then, we jointly consider factors of transmission errors, Byzantine attacks, and privacy protection to derive an upper bound on the bias of PRoBit+ aggregation, as described in Lemma \ref{lemma-6}.

    \begin{lemma}\label{lemma-5}
        In each FL iteration with lossless FedAvg aggregation on regular clients, the following is satisfied:
        \begin{align*}
            F(\bar{\bm w}_{t+1})
            \leq {} &F(\bm w_{t})-\left(\frac{1-\gamma B}{\lambda}-\frac{LB(1+\gamma)}{\bar{\lambda}\lambda} \right. \notag  \\
            & \left. -\frac{L(1+\gamma)^{2}B^{2}}{2\bar{\lambda}^{2}}\right)\times\left\|\nabla F(\bm w_{t})\right\|^{2}.
        \end{align*}
    \end{lemma}

    \begin{lemma}\label{lemma-6}
        The expected parameter deviation after PRoBit+ aggregation satisfies
        \begin{align*}
            \mathbb{E}\left[\left\|\bm \theta-\hat{\bm \theta}\right\|\right] 
             \le \|\bm \theta\|+2\sqrt{2}\beta\|\bm b\|+\sqrt{\frac{4\|\bm b\|_1}{M}\left(1+\frac{1}{\epsilon}\right)\Delta_1}. 
        \end{align*}
    \end{lemma}

    Integrating these two lemmas derives the expected loss to decrease during a single FL training iteration, as stated in Lemma \ref{theo_convergence}. 

\begin{lemma}
    In each FL iteration with PRoBit+ aggregation, the following is satisfied:
    \begin{align*}
        &\mathbb{E}\left[F(\bm w_{t+1})\right]
        \leq \mathbb{E}\left[F(\bm w_{t})\right]-\left(\frac{1-\gamma B}{\lambda} -\left(\frac{L_0}{\alpha}+\frac{L}{\lambda}\right)\right. \notag \\
        &\left.\frac{B(1+\gamma)}{\bar{\lambda}}-\frac{L(1+\gamma)^{2}B^{2}}{2\bar{\lambda}^{2}}\right)\times\left\|\nabla F(\bm w_{t})\right\|^{2} \notag \\
         & +L_0\left(2\sqrt{2}\beta\|\bm b_t\|+\sqrt{\frac{4\|\bm b_t\|_1}{M}\left(1+\frac{1}{\epsilon}\right)\Delta_1}\right).
    \end{align*}
    \label{theo_convergence}
\end{lemma}
By fixing the value of $ \bm b $, the loss reduction across iterations can be directly aggregated to establish an upper bound on the FL convergence rate. However, as training evolves, the norm of the aggregated gradients generally decreases, so a fixed $ \bm b $ would produce a progressively looser upper bound. To address this, we define a safety margin $\bm \zeta $ that measures the gap between  $ \bm b $ and the model updates in Definition \ref{margin}, resulting in a tighter upper bound on the convergence rate, as demonstrated in the theorem.
The proofs for the aforementioned lemmas and Theorem \ref{theorem-3} are provided in Appendices \ref{lemma-5-proof} and \ref{theorem-3-proof} of the supplementary material, respectively.
\end{myproof}

The convergence analysis indicates that an increase in data heterogeneity $B$ raises the upper bound of the convergence rate. 
Reducing the coefficient $\lambda$ of model regularization can mitigate this trend, thereby enhancing the FL convergence rate. This provides theoretical validation for the effectiveness of personalized FL in tackling heterogeneity.

From a security perspective, the convergence analysis also indicates that as the privacy protection requirement $\epsilon$ and the proportion of Byzantine clients $\beta$ improve, the upper bound of the convergence rate will likewise increase. 
However,  owing to PRoBit+'s estimation properties, the performance loss induced by privacy protection can be asymptotically eliminated at a rate of $\mathcal{O}(1/M)$ as the number of participating clients $M$ scales. 
This advantage originates from the ML-based aggregation method: When an increasing number of clients jointly infer the mean parameters $\bm \theta$, the PRoBit+ estimation errors diminish as  $\mathcal{O}(1/M)$ while preserving per-client randomized responses, thus guaranteeing their local DP. 
This aligns well with large-scale FL systems, indicating that PRoBit+ can preserve client privacy without significantly impacting training performance.

\subsection{Discussion on Quantization Parameter $\bm b$}

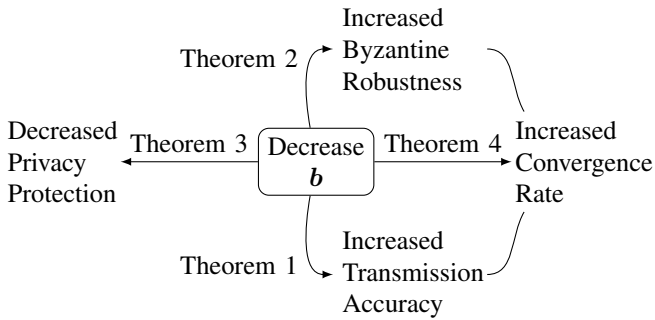
\begin{figure}[htbp]
    \begin{tikzpicture}[scale=0.5]
        \node (privacy) at (-2.7,0) {\parbox{1cm}{Decreased \\ Privacy \\ Protection}};
        \node (byzantine) at (7,3) {\parbox{1.8cm}{Increased \\ Byzantine \\Robustness}};
        \node (transmission) at (7,-3) {\parbox{1.8cm}{Increased \\ Transmission \\ Accuracy}};
        \node (convergence) at (10.8,0) {\parbox{1cm}{ Increased \\ Convergence Rate}};
        \node[draw,rounded corners] (b) at (4.5,0) {\parbox{1.3cm}{\centering Decrease \\ $\bm b$}};

        \draw[->] (b) -- +(-5.2,0)(privacy) node[midway, above, sloped] {Theorem \ref{dptheorem}};
        \draw[->] (b) -- +(5.2,0)(convergence) node[midway,above, sloped] {Theorem \ref{theorem-3}};
        \draw[->] (b) to[out=100,in=180]   node[auto] {Theorem \ref{theorem-1}} (byzantine);
        \draw[->] (b) to[out=260,in=180]   node[auto,swap] {Theorem \ref{statistic_theorem}} (transmission);
        \draw[-] (byzantine) to[out=0,in=120]   (convergence);
        \draw[-] (transmission) to[out=0,in=240]   (convergence);

    \end{tikzpicture}
    \caption{Comprehensive impacts of quantization parameter $\bm b$.}
    \label{fig:impact_of_b}
\end{figure}

We observe that the quantization parameter $\bm b$  is a key element in all PRoBit+ analyses. However, its optimization is constrained by conflicting requirements. On one hand, stronger DP necessitates greater quantization randomness, thereby requiring a larger magnitude of $\bm b$. On the other hand, enhancing Byzantine robustness and transmission accuracy—and thereby improving the FL convergence rate—necessitates a smaller magnitude of $\bm b$. This inherent conflict makes $\bm b$ a critical parameter for balancing privacy, security, and convergence rate, as illustrated in Fig.  \ref{fig:impact_of_b}. 

\section{Numerical Experiments}\label{numerical_sec}

\subsection{Simulation Setup}
We perform image classification tasks on FMNIST dataset and CIFAR-10 dataset to verify the performance of PRoBit+ and compare it with benchmark methods.
Both datasets are trained under heterogeneous data distributions.
For the FMNIST dataset, we consider an FL scenario with 100 clients, each employing a CNN. Each client randomly selects up to two classes of data, adhering to the data partitioning strategy described in \cite{McMahan2017}. In the case of the CIFAR-10 dataset, we consider an FL setup with 50 clients, each utilizing a ResNet-18 architecture. During data partitioning, each client randomly selects training samples associated with 6 different class labels.
To simplify the experiments, each local training continues for 5 epochs with a batch size of 10. The learning rate is consistently set to $\eta = 0.01$, and the penalty parameter $\lambda$ is set to $0.2$. We utilize a stochastic gradient descent (SGD) optimizer with momentum of $0.5$, and set communication rounds $T = 300$ for both datasets.
The privacy loss is set to $\epsilon = 0.1$, and the $l_1$-sensitivity is calculated as $\Delta_1=0.02\eta$ \cite{zhu2022bridging}. The simulation code is openly available at https://github.com/mh-lan/PRoBitPlus.

We compare PRoBit+ with four benchmark methods: FedAvg, FedAvg with geometric median (Fed-GM) \cite{Yin2018}, signSGD with Majority Vote (signSGD-MV) \cite{Bernstein2019} and Byzantine-Robust Stochastic Aggregation (RSA) \cite{Li2019}.
The signSGD-MV method leverages the bit robust aggregation method in \cite{Bernstein2019}, where each client uploads the signs of their model updates to the server. The server then aggregates these updates by selecting the majority sign, either $+1$ or $-1$, as the consensus value.
RSA incorporates an $l_1$ norm penalty term, $\eta\left|\bm w^m-\bm w\right|_1$, and conducts training via distributed SGD. In each communication round, each client uploads the sign of the difference between its local model and the server model, denoted as $\text{sign}(\bm w^m-\bm w)$.
The server then accumulates the received sign values and multiplies the accumulated value by an aggregation coefficient to compute the model update.
For both of these bit aggregation algorithms, the aggregation coefficient is set to $0.01$.

\subsection{Experimental Setting of $\bm b$}
\label{b-discussion}
In an ideal scenario with omniscient FL clients, the optimal value for $\bm b$ would be $ b_i  = \max_m|\delta^{m}_{i}|+\left(1+\frac{1}{\epsilon}\right)\Delta_1 $, which maximizes both Byzantine resilience and transmission precision while preserving DP. However, determining this optimal value requires a comparison of parameters from all clients, which is impossible due to the real-world communication constraints. To address this, previous studies have employed a fixed value to replace $\max_m |\delta_{m,i}|$, truncating updates that surpass this threshold \cite{Jin2024b}. Although effective in FL training, this technique significantly underperforms the optimal setting in simulations.

The performance gap between the fixed and the optimal $\bm b$ values may stem from the fact that a fixed parameter fails to adapt to the evolving  FL models. During the early training stages, truncating large updates might impede the learning process, while later, as the model approaches convergence, a fixed quantization parameter may lack the necessary granularity for model refinement, resulting in inferior final results.
To address these challenges, we propose dynamically adjusting $\bm b$ during training to enhance overall FL performance.

Conventional methods typically utilize a validation dataset to track training progress; however, hosting this dataset on the server compromises client privacy. Alternative methods that requires clients to upload their local results increases communication costs and introduces additional Byzantine vulnerabilities. To address these limitations, we propose a lightweight mechanism: each client transmits a one-bit signal to indicate whether their local loss has increased or decreased during training. The server employs a majority vote process to aggregate these signals and evaluate the overall training status. 
Specifically, each element of $\bm b$ is initially set to 0.01 and dynamically adjusted based on the training progress. When the overall training loss decreases, $b_i$ is increased by $1\%$; conversely, when the overall training loss increases, $b_i$ is decreased by $2\%$. Byzantine attacks and privacy protection are excluded to more clearly demonstrate the impact of $\bm b$.

\begin{figure}[h]
    \centering
    \includegraphics[width=0.48\textwidth]{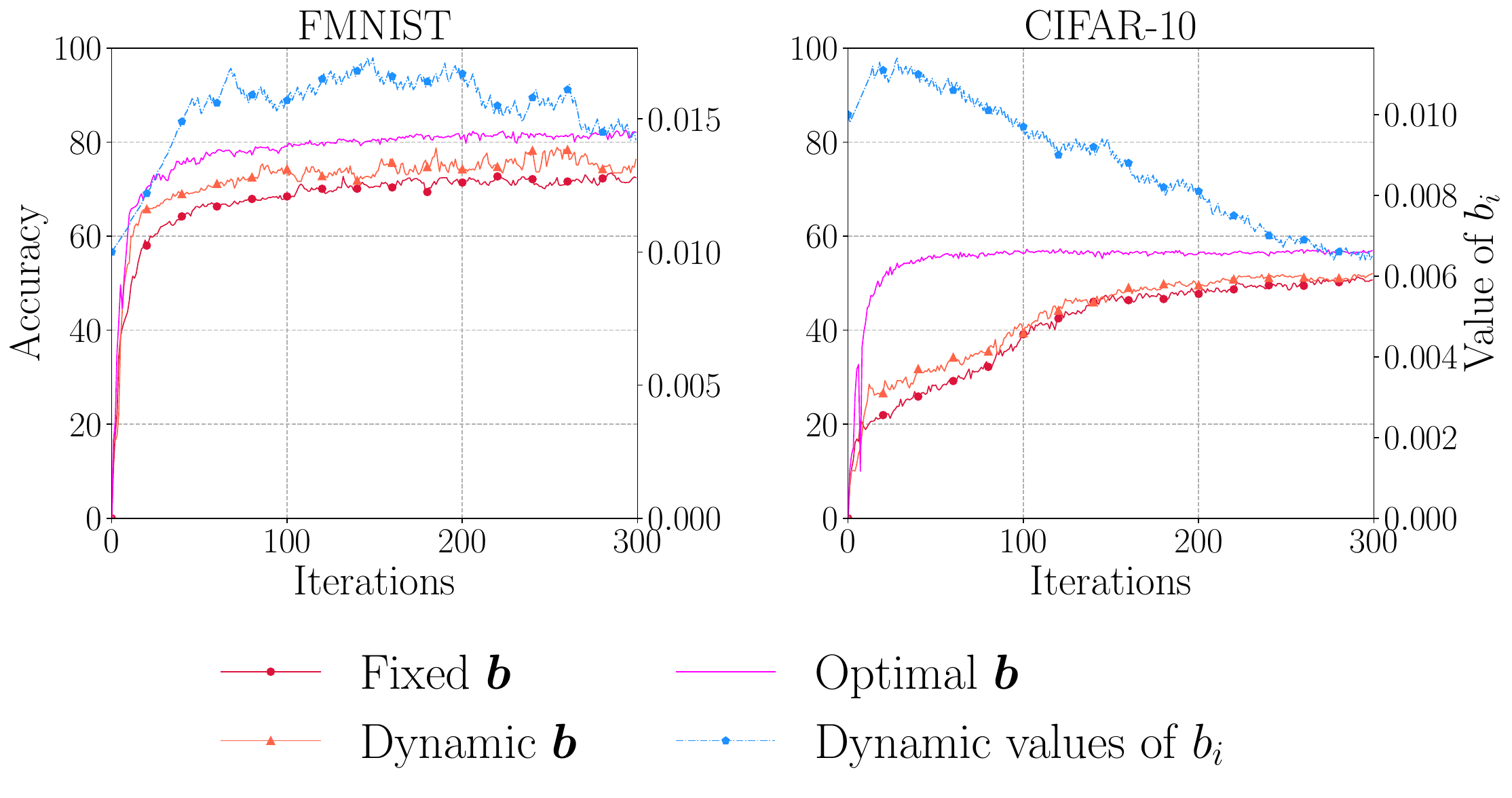}
    \caption{Training progress with different settings of $\bm b$.} \label{fig:dynamic_b}
\end{figure}

The simulation results with 20 clients training the FMNIST dataset and the CIFAR-10 dataset validate our analysis of parameter  $\bm b$, as shown in Fig.  \ref{fig:dynamic_b}. In both datasets, training results utilizing dynamic $\bm b$ exhibit higher accuracy compared to those employing a fixed $\bm b$. In the early training stages, the local loss primarily decreases with large-magnitude model updates, hence the progressively increasing $\bm{b}$ achieves a faster training rate compared to the fixed $\bm{b}$. As training progresses, the loss typically displays oscillations or an upward trend. If $\bm b$ continues to rise, quantized updates may shift towards a more uniform distribution between $1$ and $-1$, consequently escalating the loss and creating a detrimental cycle, thus hindering FL training. The dynamically adjusted $\bm b$ resolved this issue by adaptively reducing $\bm b$, thereby achieving a training performance closer to the results with the optimal $\bm b$. 
Furthermore, despite similar trends of increase and decrease in $\bm b$ on both datasets, $\bm b$ on CIFAR-10 exhibits significantly more frequent decreases than FMNIST, confirming its greater training difficulty.
Therefore, $\bm b$ can also serve as an indicator of training status.

\subsection{Communication Overhead \& Privacy Loss}

\begin{figure}[h]
    \centering
    \includegraphics[width=0.48\textwidth]{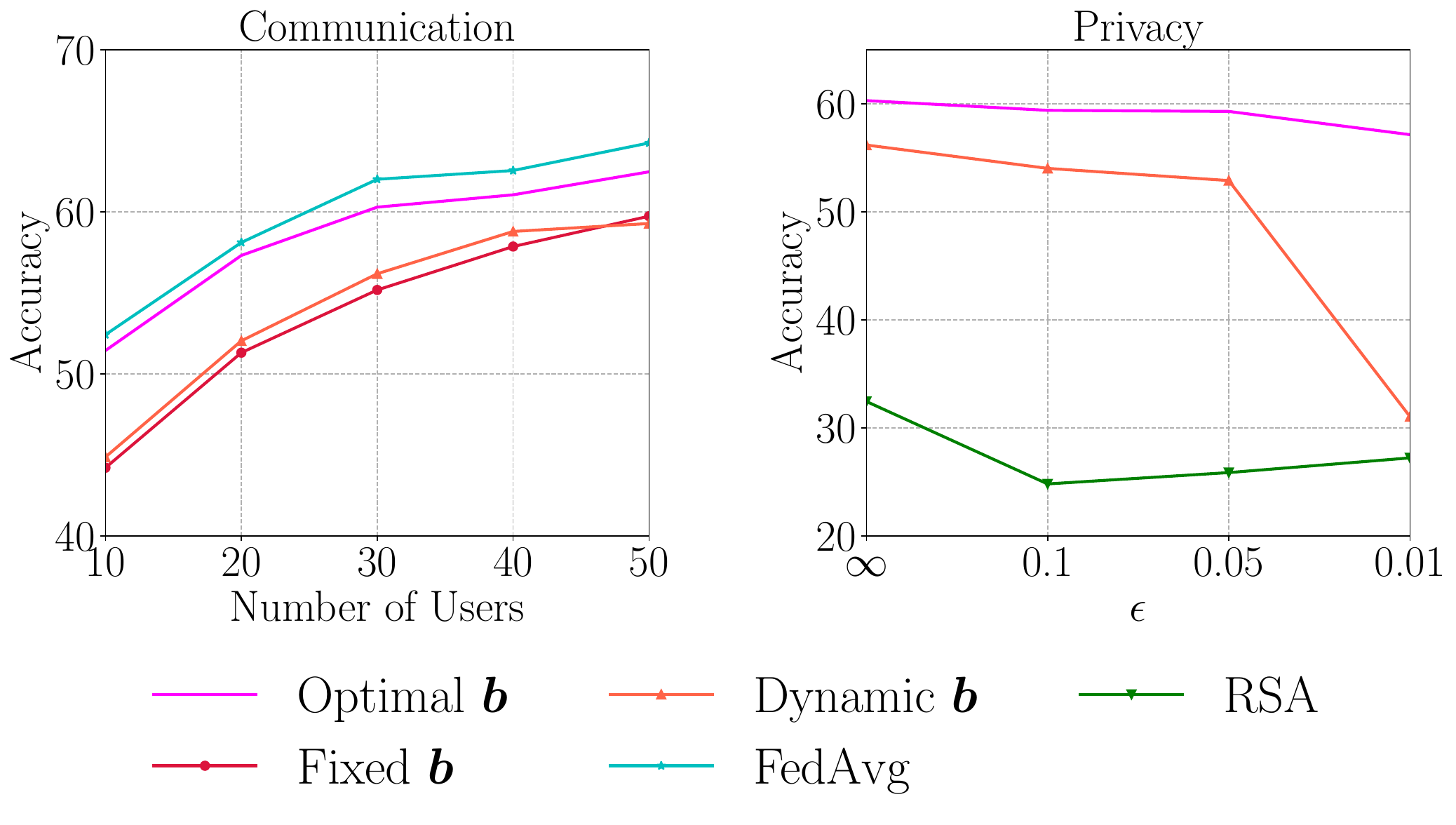}
    \caption{Test accuracy versus client numbers and privacy loss on CIFAR-10.} 
    \label{fig:commu_privacy}
\end{figure}

This subsection investigates the impact of communication overhead and privacy loss under Byzantine attack-free conditions, as depicted in Fig. \ref{fig:commu_privacy}.

The left subfigure of Fig. \ref{fig:commu_privacy} illustrates the FL test accuracy as a function of the increasing number of clients in the absence of privacy protection. It shows that the test accuracy of the proposed ML-based transmission scheme gradually enhances as the number of users grows, which corroborates the statistical properties in Theorem \ref{statistic_theorem} and the FL convergence rate in Theorem \ref{theorem-3}.  As PRoBit+ utilizes single-bit quantization for each parameter component, it reduces the communication overhead by a factor of 32 compared to the baseline FedAvg scheme employing 32-bit precision. 
When considering an FL system with $M$ users, the communication overhead of PRoBit+ amounts to merely $1/(32M)$ of FedAvg’s transmission cost.
Specifically, when $M$ increases from 10 to 50, the communication cost is reduced from 1/320 to 1/1600 times that of FedAvg, while the performance gap between PRoBit+ and FedAvg only increases from $1\%$ to $3\%$, highlighting the enhanced communication efficiency of the proposed single-bit aggregation method.

The right subfigure of Fig. \ref{fig:commu_privacy} depicts the FL test accuracy under different privacy loss $\epsilon$, with a fixed client number of 30. 
In accordance with the convergence analysis in Theorem \ref{theorem-3}, which indicates that the negative impact of privacy protection diminishes asymptotically at a rate of $\mathcal{O}(1/M)$, the proposed PRoBit+ demonstrates negligible performance degradation when $\epsilon \leq 0.05$ in comparison to the RSA algorithm. However, a further decrease of $\epsilon$ to $0.01$ leads to significant performance deterioration, which is consistent with the impacts of the quantization parameter $\bm b$ discussed in Fig. \ref{fig:impact_of_b}.

\begin{figure*}[h]
    \centering
    \includegraphics[width=\textwidth]{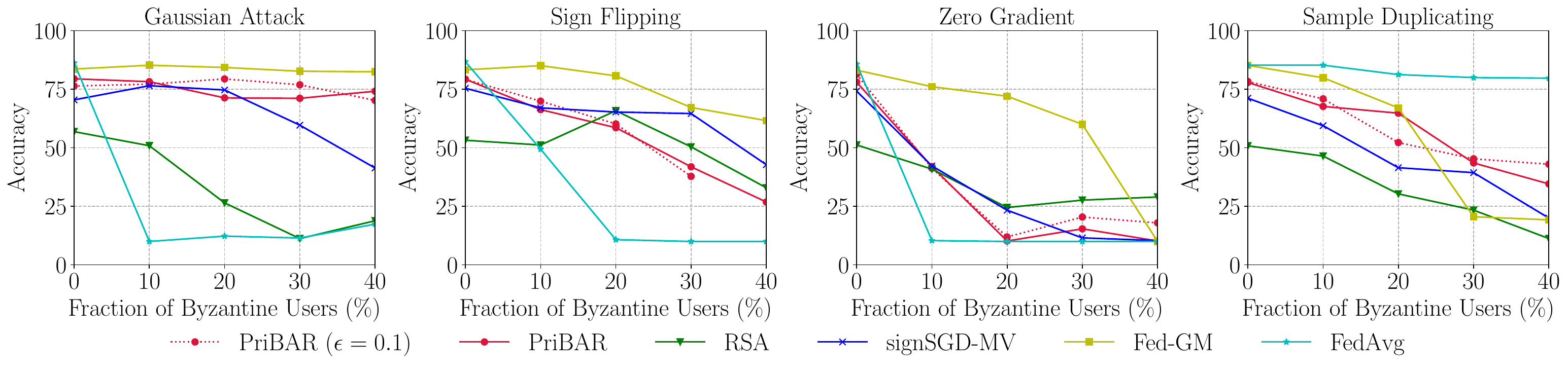}
    \caption{Test accuracy versus proportions of Byzantine clients on FMNIST.} \label{fig:fmnist_frac}
\end{figure*}

\begin{figure*}[h]
    \centering
    \includegraphics[width=\textwidth]{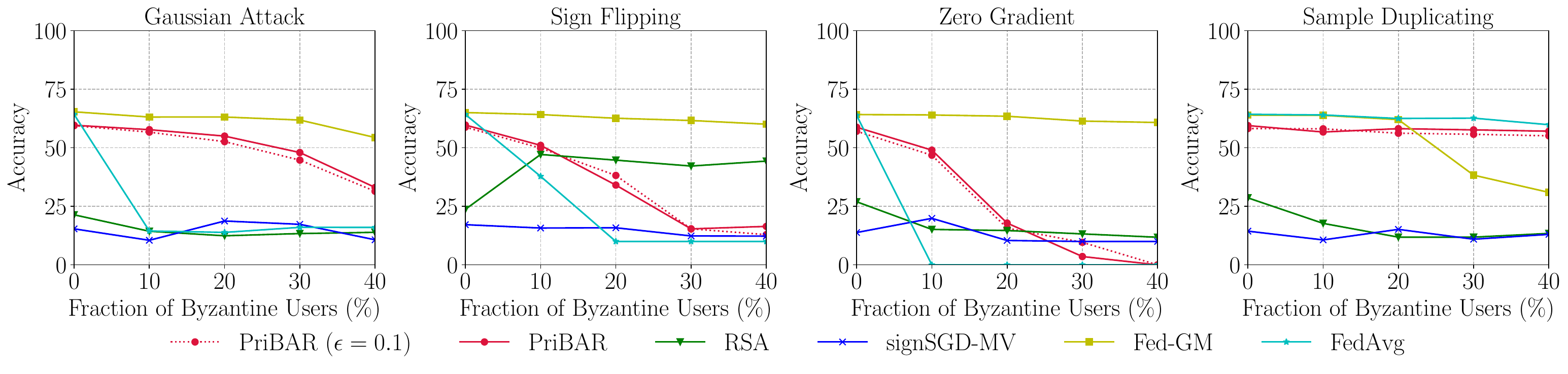}
    \caption{Test accuracy versus proportions of Byzantine clients on CIFAR-10.} \label{fig:cifar_frac}
\end{figure*}

\subsection{Byzantine Robustness}
We examine the Byzantine robustness against four types of malicious updates: 1) Gaussian attack: Each Byzantine client generates a Gaussian random variable as a malicious upload value, with each component being independently and identically distributed according to $\mathcal{N}\left(0,100\right)$; 2) Sign flipping: Each Byzantine client scales its honest model update by a factor of $-5$; 3) Zero gradient: All Byzantine clients conspire to generate a same malicious value, resulting in the parameter updates that sums to zero; 
4) Sample duplicating: Each Byzantine client replicates the model update of the first honest client. 

This subsection begins with a comparison of test accuracy across varying proportions of Byzantine clients.
Then the scenario with $10\%$ Byzantine clients is selected to investigate the training stability of each algorithm in more detail.
Note that $\bm b$ is fixed as $\bm {0.01}$ in this subsection to prevent Byzantine attacks on the loss messages.

\subsubsection{FMNIST}
We first analyze the simulation results on the FMNIST dataset, as illustrated in Fig \ref{fig:fmnist_frac}. 
From the perspective of DP protection, with the number of training clients reaching 100, Theorem \ref{theorem-3} indicates that the training disruption caused by privacy protection can be effectively mitigated. The simulation results confirm the conclusion in multiple training configurations, indicating that the implementation of privacy protection only causes minor fluctuations in test accuracy.
Furthermore, when all clients are honest, the proposed PRoBit+ algorithm outperforms other bit transmission algorithms with equivalent communication overhead, even approaching the accuracy of FedAvg algorithm utilizing full-precision transmission, realizing our design vision.

However, this design principle becomes a double-edged sword when defending against Byzantine attacks. 
Given the averaging nature of FedAvg, zero-mean noise, such as Gaussian noise, is expected to exhibit a degree of self-cancellation, as positive and negative deviations are equally probable.
Combined with the bit amplitude restriction on model updates, PRoBit+ exhibits comparable robustness to Fed-GM, a full-precision defense scheme with computational complexity as high as $\mathcal{O}\left(M^2\right)$, when defending against Gaussian attacks.
However, this concurrently results in PRoBit+'s suboptimal defense performance against collusive zero gradient attacks, which is marginally inferior to current bit-based Byzantine defense strategies.

The stochastic quantization based on parameter amplitudes also presents both advantages and disadvantages. 
When dealing with sample duplicating attacks, deterministic data processing will lead aggregation results to diverge, thereby impeding the training process.
In contrast, PRoBit+, by introducing randomization, mitigates the influence of individual client parameters during aggregation, yielding better Byzantine robustness than signSGD-MV and even outperforming Fed-GM when Byzantine proportions exceeding $30\%$.
But at the same time, this characteristic also makes PRoBit+ less effective when Byzantine clients showcase particularly large amplitudes, resulting in inferior defense performance compared to existing bit-based Byzantine defense methods against sign flipping attacks.

\subsubsection{CIFAR-10}
The ResNet-18 architecture is employed to train the CIFAR-10 dataset, with the results illustrated in Fig \ref{fig:cifar_frac}.  
The FL implement of deep networks requires improved parameter transmission accuracy, leading to the non-convergence of signSGD-MV method in all configurations. 
The performance of RSA has likewise declined dramatically as a result of increased dataset heterogeneity. 
PRoBit+, however, maintains the desirable properties observed in training the FMNIST dataset, effectively mitigating the performance loss caused by DP, and achieving FedAvg-like performance in the absence of Byzantine attacks.
In terms of Byzantine robustness, PRoBit+ shows similar characteristics on the CIFAR-10 dataset as on the FMNIST dataset. Due to its superiority in training deep networks, PRoBit+ outperforms other bit-based robust algorithms.

\begin{figure*}[h]
    \centering
    \includegraphics[width=\textwidth]{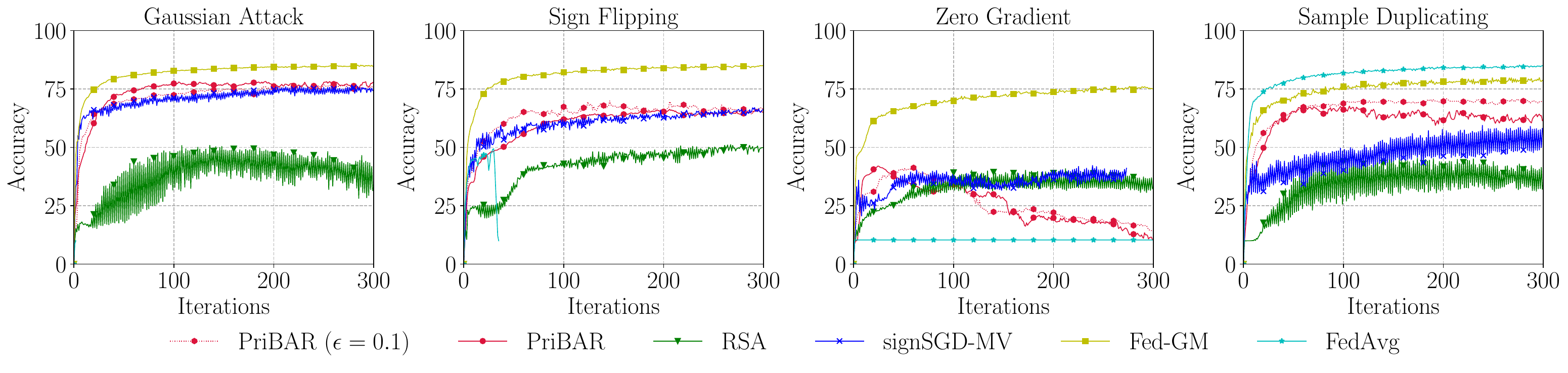}
    \caption{Test accuracy versus training iterations with $10\%$ Byzantine proportions on FMNIST.} \label{fig:fmnist_performance}
\end{figure*}

\begin{figure*}[h]
    \centering
    \includegraphics[width=\textwidth]{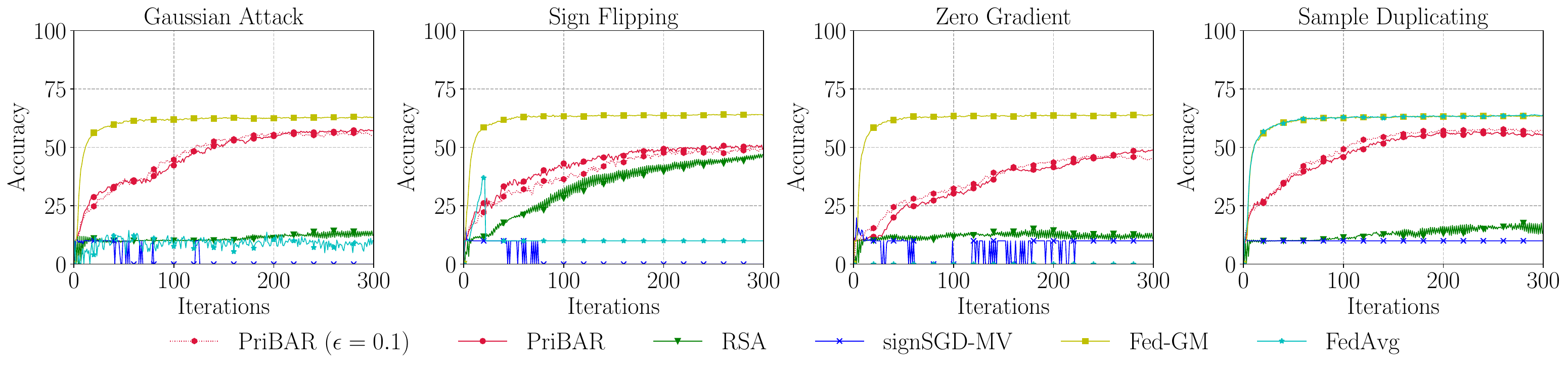}
    \caption{Test accuracy versus training iterations with $10\%$ Byzantine proportions on CIFAR-10.} \label{fig:cifar_performance}
\end{figure*}

\subsubsection{Case study with $\beta = 10\%$}

We analyze the FL training stability with $10\%$ Byzantine proportions. Fig. \ref{fig:fmnist_performance} and Fig. \ref{fig:cifar_performance} illustrate the training process, whereas Table \ref{table_simu} presents the final test accuracies.
Except for its defensive performance against zero gradient attacks on the FMNIST dataset, PRoBit+ demonstrates stable learning with minimal training fluctuations in most scenarios. 
The RSA and signSGD-MV algorithms, both integrating bit transmission aggregation, exhibit significant fluctuations throughout the training process, indicating a dependence on the specified training step size. 
Furthermore, despite the increased training difficulty posed by CIFAR-10 compared to FMNIST, PRoBit+ exhibits robust stability against all tested attacks, in contrast to other bit transmission algorithms that generally fail to converge.
It even approaches the performance of Fed-GM under Gaussian attacks and sample replication attacks, indicating the algorithm's strong adaptability in the FL of deep networks.

\begin{table*}[!ht]
    \centering
    \caption{Test accuracy of each method with $10\%$ Byzantine proportions.}
    \label{table_simu}
    \begin{tabular}{|l|c|c|c|c|c|c|}
    \hline
    \multicolumn{7}{|c|}{FMNIST}\\ \hline
        ~ &\textbf{PRoBit+} & \textbf{PRoBit+} ($\epsilon=0.1$) & \textbf{RSA} & \textbf{signSGD MV} & \textbf{Fed GM} & \textbf{FedAvg}  \\ \hline
        \textbf{Gaussian attack} & 78.24 & 77.19 & 50.89 & 76.51 & 85.27 & 10  \\ \hline
       \textbf{Sign flipping }& 66.31 & 69.92 & 51.19 & 67.05 & 85.1 & 49.42  \\ \hline
        \textbf{Zero gradient} & 42.01 & 41.28 & 40.9 & 42.27 & 76.11 & 10.36  \\ \hline
        \textbf{Sample duplicating} & 67.68 & 70.91 & 46.47 & 59.45 & 79.91 & 85.3  \\ \hline
        \multicolumn{7}{|c|}{CIFAR-10}\\ \hline
        ~ & \textbf{PRoBit+} & \textbf{PRoBit+}($\epsilon=0.1$) & \textbf{RSA} & \textbf{signSGD MV} & \textbf{Fed GM} & \textbf{FedAvg}  \\ \hline
        \textbf{Gaussian attack} & 57.74 & 56.69 & 14.35 & 10.5 & 63.14 & 14.47  \\ \hline
        \textbf{Sign flipping} & 51.05 & 49.87 & 47.15 & 15.73 & 64.19 & 37.8  \\ \hline
        \textbf{Zero gradient} & 49 & 46.82 & 15.14 & 19.87 & 64.02 & 0  \\ \hline
        \textbf{Sample duplicating} & 56.74 & 58.11 & 17.72 & 10.66 & 63.85 & 64.03  \\ \hline
    \end{tabular}
\end{table*}

\section{Conclusion}\label{conclusion_sec}
This paper proposes a PRoBit+ aggregation method to jointly address communication overhead, Byzantine attacks, and privacy leakage issues in heterogeneous FL. PRoBit+ applies magnitude-based stochastic one-bit quantization to model updates, significantly reducing communication costs while preventing malicious clients from uploading drastically deviated updates, thus achieving Byzantine robustness. By controlling the randomness of quantization, we prove that PRoBit+'s inherent stochastic mechanism satisfies $(\epsilon, 0)$-DP. Additionally, we propose an ML estimation based parameter aggregation method on the server side, which not only achieves unbiased estimation of aggregated parameters but also enables dynamic adjustment of model update step sizes, thereby eliminating the dependence on manually set training step sizes.

Considering a personalized FL framework with model regularization in heterogeneous scenarios, we theoretically derive a convergence rate upper bound on the PRoBit+ based FL. 
This bound explicitly reveals the interplay of data heterogeneity, security level and transmission accuracy on training rate, as well as how to make targeted adjustments through the quantization randomness parameter $\bm b$. Moreover, the upper bound indicates that the impact of privacy protection and one-bit transmission errors on the convergence rate can be asymptotically mitigated at a rate of $\mathcal{O}(1/M)$ as the number of FL clients $M$ increases.

In experiments, we first design a dynamic adjustment method for the quantization parameter $\bm b$, resulting in an accelerated training rate and improved test accuracy compared to current methods using a fixed $\bm b$. 
We further compare PRoBit+ experimentally against several benchmark methods under various Byzantine attacks.
When training ResNet-18, PRoBit+ achieves superior accuracy compared to existing bit-transmission-based robust aggregation methods, even outperforming median-based methods under sample duplication attacks.
The results also show that without Byzantine attacks, one-bit PRoBit+ achieves comparable test accuracy to full-precision FedAvg.

This paper advances the understanding of bit aggregation in FL and offers insights for the secure and effective deployment of deep networks in large-scale FL systems.
Future work will explore position-aware quantization strategies, enabling finer-grained control over privacy and convergence. 
We will also explore personalized FL through partial network updates to further enhance Byzantine robustness and reduce communication overhead.

\bibliographystyle{IEEEtran}
\bibliography{ref}

\clearpage
\setcounter{page}{1}
\twocolumn[
    \begin{@twocolumnfalse}
        \section*{\Large \centering{Supplementary Material: One-Bit Model Aggregation for Differentially Private and Byzantine-Robust Personalized Federated Learning
          }}
          \vspace{5pt}
          \centerline{\large Muhang Lan, Song Xiao, and Wenyi Zhang
          }
          \centerline{ }
    \end{@twocolumnfalse}
]

{\appendices
\section{Proof of Theorem \ref{statistic_theorem}}\label{statistic_theorem-proof}
\begin{myproof}[Proof]
    We prove each part of Theorem \ref{statistic_theorem} as follows:
    \begin{enumerate}
        \item  We first consider the estimation of the parameter $\bm \theta$ in the $i$-th dimension, i.e., $\theta_i$.
        Taking the logarithm of the joint probability function \eqref{probability_func} yields
            \begin{align*}
            & \log \mathbb{P}(c^1_i,c^2_i,\cdots,c^M_i;\theta_i)  \\
             = {} & N_i\left(\frac{1}{2}+\frac{\theta_i}{2b}\right)+(M-N_i)\left(\frac{1}{2}-\frac{\theta_i}{2b}\right) \\
             = {} & \frac{1}{2}\left(M\left(1-\frac{\theta_i}{b}\right)+N\frac{2\theta_i}{b}\right)                     \\
             = {} & \frac{1}{2}\left(M+\frac{\theta_i}{b}\left(2N_i-M\right)\right)                                     \\
             = {} & \frac{1}{2}\left(M+\frac{M}{b^2}\theta_i \hat{\theta}_i \right).
            \end{align*}

        By the Factorization Theorem, it can be concluded that the statistic $\hat{\theta}_i$ is a sufficient statistic for $\theta_i$. Since different components are independent of each other, $\hat{\bm{\theta}}$ is a sufficient statistic for $\bm{\theta}$.
        \item The expectation of $\hat{\theta}_{i}$ is calculated as
        \begin{align}
            \mathbb{E}\left[\hat{\theta}_{i}\right]&= \frac{2\mathbb{E}\left[N_i\right]-M}{M}b_i \notag \\
            &= \frac{2M\mathbb{P}(c^m_i=1)-M}{M}b_i \notag \\
            &= \left(2\left(\frac{1}{2}+\frac{1}{2b_i}\theta_{i}\right)-1\right)b_i \label{eq3-22}\\
            &= \theta_i, \notag 
        \end{align}
        where \eqref{eq3-22} comes from \eqref{eq3-6}.
        \item We first need to calculate two key statistics, including
        \begin{align*}
            \mathbb{E}\left[N_i\right]
             & = \sum_{m=1}^{M}\mathbb{P}\left(c_i^m=1\right)                                   \\
             & = \sum_{m=1}^{M}\mathbb{E}\left[\mathbb{P}\left(c_i^m=1|\delta_i^m\right)\right] \\
             & =\sum_{m=1}^{M}\mathbb{E}\left[\frac{b_i+\delta_i^m}{2b_i}\right]                \\
             & =\frac{M}{2}\left(1+\frac{\mathbb{E}\left[\delta_i^m\right]}{b_i}\right)         \\
             & =\frac{M}{2}\left(1+\frac{\theta_i}{b_i}\right)
        \end{align*}
    
        \begin{align*}
                & \mathbb{E}\left[N_i^2\right] \notag \\

            = {} & \mathbb{E}\left[\sum_{m=1}^{M} \mathbb{I}^2\left\{c_i^m=1\right\} \right. \notag \\
            & + \left.\sum_{i\ne j}\mathbb{I}\left\{c_i^m=1\right\}\mathbb{I}\left\{c_j^m=1\right\}\right]                    \\
            = {} & \sum_{m=1}^{M}\mathbb{E}\left[\mathbb{I}\left\{c_i^m=1\right\}\right] \notag \\
            &+\sum_{i\ne j}\mathbb{E}\left[\mathbb{I}\left\{c_i^m=1\right\}\mathbb{I}\left\{c_j^m=1\right\}\right] \\
            = {} & M \mathbb{P}\left(c_i^m=1\right) +M(M-1)\mathbb{E}^2\left[\mathbb{I}\left\{c_i^m=1\right\}\right]                                                                         \\
            = {} & M\left(\frac{b_i+\theta_i}{2b_i}\right)+M(M-1)\left(\frac{b_i+\theta_i}{2b_i}\right)^2                                                                            \\
            = {} & M\left(\frac{b_i+\theta_i}{2b_i}\right)\left(1+\left(M-1\right)\frac{b_i+\theta_i}{2b_i}\right)                                                                          \\
            = {} & M\left(\frac{b_i+\theta_i}{2b_i}\right)\left(\frac{\left(M+1\right)b_i+\left(M-1\right)\theta_i}{2b_i}\right).        
        \end{align*}
    
        By calculating the variance, we get
        \begin{align}
            & \mathbb{E}\left[\left(\theta_{i}-\hat{\theta}_{i}\right)^2\right] \notag \\
            = {} & \mathbb{E}\left[\frac{2N_i-M}{M}b_i\right]^2-\theta_{i}^2                                                     \notag \\
            = {} & \left(\frac{b_i}{M}\right)^2\mathbb{E}\left[4N_i^2-4N_iM+M^2\right]-\theta_{i}^2                             \notag \\
            = {} & \left(\frac{b_i}{M}\right)^2\cdot \left(4\mathbb{E}\left[N_i^2\right]-4M\mathbb{E}\left[N_i\right]+M^2\right)
            -\theta_{i}^2                                                                                                     \notag \\
            = {} & \frac{b_i^2-\theta_i^2}{M}. \label{eq3-37}
        \end{align}
        From this, we can obtain the transmission error as
        \begin{align*}
            \mathbb{E}\left[\left\|\bm{\theta}-\hat{\bm{\theta}}\right\|^2\right]
             & = \mathbb{E}\left[\sum_{i=1}^{d}\left(\theta_{i}-\hat{\theta}_{i}\right)^2\right] \\
             & = \frac{\sum_{i=1}^{d}\left(b_i^2-\theta_i^2\right)}{M}, 
        \end{align*}
        where $d$ is the length of the vector.
    \end{enumerate}
   
\end{myproof}

\section{Proofs of Lemma \ref{lemma-5}, Lemma \ref{lemma-6} and Lemma \ref{theo_convergence}}\label{lemma-5-proof}
\begin{myproof}[Proof of Lemma \ref{lemma-5}]
    We define $\bm e^{m}_{t+1}= \nabla f_{m}(\bm w^{m}_{t+1})+\lambda(\bm w^{m}_{t+1}-\bm w_{t})$, then the client $m$'s local training  satisfies
    \begin{align}
        \|\bm e^{m}_{t+1}\| & \leq\gamma\|\nabla f_{m}(\bm w_{t})\|, \label{eq3-66}
    \end{align}
   which comes from the $\gamma$-inexact solution in Definition \ref{inexact}.
    Define $\bar{\bm w}_{t+1}=\mathbb{E}_{m}\left[\bm w^{m}_{t+1}\right]= \frac{1}{R}\sum_{m=1}^{R}  \bm w^m_{t+1}$, we get
    \begin{align*}
        \bar{\bm w}_{t+1}-\bm w_t=\frac{-1}{\lambda}\mathbb{E}_m\left[\nabla f_m(\bm w^m_{t+1})\right]+\frac{1}{\lambda}\mathbb{E}_m\left[\bm e^m_{t+1}\right].
    \end{align*}
    With $\bar{\lambda}=\lambda-L_{-}>0$, we obtain
    \begin{align}
        \nabla^2h_m-\bar{\lambda}\bm I
         & =  \nabla^2f_m+\lambda \bm I -(\lambda-L_{-})\bm I \notag \\
         & =  \nabla^2f_m+L_{-}\bm I                         \notag  \\
         & \succeq \bm 0, \label{eq3-67}
    \end{align}
    where \eqref{eq3-67} comes from the Assumption \ref{assup_smoothness} and shows that $h_m$ is $\bar \lambda$-strongly convex.
    Further define $\tilde{\bm w}^m_{t+1}=\arg\min_{\bm w}h_m(\bm w;{\bm w}_t)$, then we get
    \begin{align}
        \bar{\lambda}\|\tilde{\bm w}^m_{t+1}-\bm w^m_{t+1}\|
         & \le \|\nabla h(\tilde{\bm w}^m_{t+1})-\nabla h({\bm w}^m_{t+1}) \| \notag \\
         & = \|\nabla h({\bm w}^m_{t+1}) \|                                  \notag  \\
         & = \|\bm e^{m}_{t+1}\|                                         \notag      \\
         & \le \gamma\|\nabla f_{m}(\bm w_{t})\|. \label{3-70}
    \end{align}

Similarly, using the $\bar{\lambda}$-strong convexity of $h_m$, we obtain
\begin{align}
    \bar{\lambda}\|\tilde{\bm w}^m_{t+1}-\bm w_{t}\|
    & \le \|\nabla h_m({\bm w}_{t}) \|                               \notag  \\
    & = \| \nabla f_m({\bm w}_{t})+\lambda({\bm w}_{t}-{\bm w}_{t})\| \notag \\
    & = \| \nabla f_m({\bm w}_{t})\|. \label{3-73}
\end{align}
By applying the triangle inequality to \eqref{3-70} and \eqref{3-73}, we get
\begin{align}
    \|\bm w^m_{t+1}-\bm w_t\|\leq\frac{1+\gamma}{\bar{\lambda}}\|\nabla f_m(\bm w_t)\|. \label{eq3-79}
\end{align}
From this, we can derive
\begin{align}
    \|\bar{\bm w}_{t+1}-\bm w_t\|
    & \leq\mathbb{E}_m[\|\bm w^m_{t+1}-\bm w_t\|]  \label{eq3-80}                                    \\
    & \leq\frac{1+\gamma}{\bar{\lambda}}\mathbb{E}_m[\|\nabla f_m(\bm w_t)\|]      \notag     \\
    & \leq\frac{1+\gamma}{\bar{\lambda}}\sqrt{\mathbb{E}_m[\|\nabla f_m(\bm w_t)\|^2]} \label{eq3-82} \\
    & \leq\frac{B(1+\gamma)}{\bar{\lambda}}\|\nabla F(\bm w_t)\|, \label{eq3-83}
\end{align}
where \eqref{eq3-80} follows from Jensen's inequality; \eqref{eq3-82} follows from the inequality $\sqrt{\mathbb{E}\left[A^2\right]}\ge \mathbb{E}\left[A\right]$; \eqref{eq3-83} follows from Assumption \ref{assup_dissimilarity}.

Further define $\bm G_{t+1}$ to describe $\bar{\bm w}_{t+1}-\bm w_{t}=\frac{-1}{\lambda}\left(\nabla F(\bm w_{t})+\bm G_{t+1}\right)$, then we have
\begin{align*}
    \bm G_{t+1}=\mathbb{E}_m\left[\nabla f_m(\bm w^m_{t+1})-\nabla f_m(\bm w_t)-\bm e^m_{t+1}\right],
\end{align*}
whose upper bound satisfies
\begin{align}
     \|\bm G_{t+1}\| 
     \le {} & \mathbb{E}_m \left\|\left[\nabla f_m(\bm w^m_{t+1})-\nabla f_m(\bm w_t)-\bm e^m_{t+1}\right] \right\|                     \notag   \\
     \le {} & \mathbb{E}_m\left[ \left\|\nabla f_m(\bm w^m_{t+1})-\nabla f_m(\bm w_t)\|+\|\bm e^m_{t+1} \right\|\right]             \notag       \\
     \le {} & \mathbb{E}_m\left[L\|\bm w^m_{t+1}-\bm w^m_t\|+\|\bm e^m_{t+1}\|\right]                                                   \notag   \\
     \le {} & \left(\frac{L(1+\gamma)}{\bar{\lambda}}+\gamma\right)\times\mathbb{E}_m\left[\|\nabla f_m(\bm w_t)\|\right]  \label{eq3-88} \\
     \le {} & \left(\frac{L(1+\gamma)}{\bar{\lambda}}+\gamma\right)B\|\nabla F(\bm w_t)\|, \label{eq3-89}
\end{align}
where \eqref{eq3-88} follows from \eqref{eq3-79} and \eqref{eq3-66}, and \eqref{eq3-89} is derived in the same manner as \eqref{eq3-83}.

Since $ F $ is a convex combination of the local loss functions $ f_m $, it is also $ L $-Lipschitz smooth, which gives us
\begin{align}
    F(\bar{\bm w}_{t+1})
    \leq {} & F(\bm w_{t})+\langle\nabla F(\bm w_{t}),\bar{\bm w}_{t+1}-\bm w_{t}\rangle \notag  \\
    &+\frac{L}{2}\|\bar{\bm w}_{t+1}-\bm w_{t}\|^{2}   \label{eq3-94}                                                                                \\
    \leq {} & F(\bm w_{t})-\frac1\lambda\|\nabla F(\bm w_{t})\|^{2}-\frac1\lambda\langle\nabla F(\bm w_{t}),\bm G_{t+1}\rangle \notag  \\ 
    &+\frac{L(1+\gamma)^{2}B^{2}}{2\bar{\lambda}^{2}}\|\nabla F(\bm w_{t})\|^{2}                \label{eq3-95} \\
    \leq {} & F(\bm w_{t})-\left(\frac{1-\gamma B}{\lambda}-\frac{LB(1+\gamma)}{\bar{\lambda}\lambda}\right. \notag \\
    & \left. -\frac{L(1+\gamma)^{2}B^{2}}{2\bar{\lambda}^{2}}\right)\times\left\|\nabla F(\bm w_{t})\right\|^{2}, \label{eq3-96}
\end{align}
where substituting the definition of $\bm G_{t+1}$ and \eqref{eq3-83} into \eqref{eq3-94} yields \eqref{eq3-95}, and \eqref{eq3-96} follows from the Cauchy-Schwarz inequality, i.e.,
\begin{align*}
    & -\langle\nabla F(\bm w_{t}),\bm G_{t+1}\rangle \notag \\
    \le {} & \left\|\langle\nabla F(\bm w_{t}),\bm G_{t+1}\rangle \right\|                                                    \\
    \le {} & \left\|\nabla F(\bm w_{t})\right\|\left\|\bm G_{t+1}\right\|                                                    \\
    \le {} & \left\|\nabla F(\bm w_{t})\right\| \left(\frac{L(1+\gamma)}{\bar{\lambda}}+\gamma\right)B\|\nabla F(\bm w_t)\| \\
    = {} & \left(\frac{L(1+\gamma)}{\bar{\lambda}}+\gamma\right)B\|\nabla F(\bm w_t)\|^2.
\end{align*}
\end{myproof}

\begin{myproof}[Proof of Lemma \ref{lemma-6}]
    For simplicity, we omit the superscript $t$ in the following analysis. The $i$-th component of $\hat{\bm \theta}$ satisfies
\begin{align*}
    \hat{\theta}_i
    = {} & \frac{2\left(\sum_{m\in \mathcal{R}}\mathbb{I}\left\{c^m_{i}=1\right\}+\sum_{m\in \mathcal{B}}\mathbb{I}\left\{z^m_{i}=1\right\}\right)-M}{M}b_i                                                                \\
    = {} & \frac{1}{M} \left(2\left(\sum_{m=1}^{M}\mathbb{I}\left\{c^m_{i}=1\right\}+\sum_{i\in \mathcal{B}}\mathbb{I}\left\{z^m_{i}>c^m_{i}\right\} \right.\right. \notag\\
    &- \left.\left. \sum_{i\in \mathcal{B}}\mathbb{I}\left\{z^m_{i}<c^m_{i}\right\}\right)-M\right)b_i,
\end{align*}
where $c^m_{i}$ represents honest quantization results, and $z^m_i$ represents malicious messages, which can take 0 or 1 arbitrarily.

Since the server is unaware of the attacks, it is reasonable to assume that the aggregated values remain unbiased, and the deviation caused by Byzantine attacks satisfies
\begin{align*}
    & \mathbb{E}\left[ \left(\theta_i-\hat{\theta_i}\right)^2\right] \notag \\
    = {} & \mathbb{E}\left[\frac{1}{M} \left(2 \left(\sum_{m=1}^{M}\mathbb{I}\left\{c^m_{i}=1\right\}+\sum_{i\in \mathcal{B}}\mathbb{I}\left\{z^m_{i}>c^m_{i}\right\} \right.\right.\right. \notag \\
    &-\left.\left.\left. \sum_{i\in \mathcal{B}}\mathbb{I}\left\{z^m_{i}<c^m_{i}\right\}\right)-M\right)b_i\right]^2-{\theta_i}^2               \\
    = {} & \mathbb{E}\left[\frac{2\sum_{m=1}^{M}\mathbb{I}\left\{c^m_{i}=1\right\}-M}{M}b_i \right. \notag \\
    & +\left. \frac{2 \left(\sum_{i\in \mathcal{B}}\mathbb{I}\left\{z^m_{i}>c^m_{i}\right\}-\sum_{i\in \mathcal{B}}\mathbb{I}\left\{z^m_{i}<c^m_{i}\right\}\right)}{M}b_i\right]^2 \notag \\
    &-{\theta_i}^2 \\
    \le {} & 2\mathbb{E}\left[\frac{2 \sum_{m=1}^{M}\mathbb{I}\left\{c^m_{i}=1\right\}-M}{M}b_i\right]^2             \notag                                                                                                                                                         \\
    & + 2\mathbb{E}\left[\frac{2}{M} \left(\sum_{i\in \mathcal{B}}\mathbb{I}\left\{z^m_{i}>c^m_{i}\right\} \right.\right. \notag \\
    &\left.\left.-\sum_{i\in \mathcal{B}}\mathbb{I}\left\{z^m_{i}<c^m_{i}\right\}\right)b_i\right]^2-{\theta_i}^2                                                                 \\
    \le {} & 2\left(\mathbb{E}\left[\frac{2 \sum_{m=1}^{M}\mathbb{I}\left\{c^m_{i}=1\right\}-M}{M}b_i\right]^2-{\theta_i}^2\right) \notag \\
    &+2\mathbb{E}\left[2\beta b_i\right]^2+{\theta_i}^2                                                                                          \\
    \le {} & \frac{2\left(b_i^2-\theta_i^2\right)}{M}+8\beta^2 b_i^2+{\theta_i}^2,              
\end{align*}
where the last inequality follows from the analysis of parameter estimation in \eqref{eq3-37}.

Further extrapolating the result to a vector notation, we have
\begin{align}
    & \mathbb{E}\left[\left\|\bm \theta-\hat{\bm \theta}\right\|\right] \notag \\
    \le {} & \sqrt{\mathbb{E}\left[\sum_{i=1}^{d}\left(\theta_i-\hat{\theta}_i\right)^2\right]}             \label{eq3-106}                 \\
    \le {} & \sqrt{\sum_{i=1}^{d} \frac{2\left(b_i^2-\theta_i^2\right)}{M}+8\beta^2 b_i^2+{\theta_i}^2}                              \notag        \\
    \le {} & \sqrt{\|\bm \theta\|^2+8\beta^2\|\bm b\|^2+\frac{2}{M}\left(\|\bm b\|^2-\|\bm \theta\|^2\right)}                       \notag        \\
    \le {} & \|\bm \theta\|+2\sqrt{2}\beta\|\bm b\|+\sqrt{\frac{2}{M}\left(\|\bm b\|^2-\|\bm \theta\|^2\right)}            \label{eq3-104}  \\
    \le {} & \|\bm \theta\|+2\sqrt{2}\beta\|\bm b\|+\sqrt{\frac{4\|\bm b\|_1}{M}\left(1+\frac{1}{\epsilon}\right)\Delta_1},  \label{eq3-105}
\end{align}
where \eqref{eq3-106} follows from the Jensen's inequality; \eqref{eq3-104} follows from the Cauchy-Schwarz inequality; \eqref{eq3-105} follows from the DP requirement in Theorem \ref{dptheorem}, i.e.,
\begin{align*}
    \|\bm b\|^2-\|\bm \theta\|^2
    & = \sum_{i=1}^{d}\left( b_i+\theta_i \right)\left(b_i-\theta_i\right) \\
    & \le 2\left(1+\frac{1}{\epsilon}\right)\Delta_1\sum_{i=1}^{d}b_i      \\
    & = 2\left(1+\frac{1}{\epsilon}\right)\Delta_1 \|\bm b\|_1.
\end{align*}
\end{myproof}

\begin{myproof}[Proof of Lemma \ref{theo_convergence}]
    We utilize Lemma \ref{lemma-5} and Lemma \ref{lemma-6} to prove Lemma \ref{theo_convergence}. 
    Lemma \ref{lemma-5} derives an upper bound of FL convergence rate with lossless FedAvg aggregation on regular clients. However, under the influence of the PRoBit+ transmission mechanism, we need to further consider the impact of quantized transmission and Byzantine attacks.
The quantized aggregated parameter $\bm w_{t+1}$ and the lossless aggregated parameter $\bar{\bm w}_{t+1}$ satisfy
\begin{align}
    & F(\bm w_{t+1})-F(\bar{\bm w}_{t+1}) \notag \\
    \leq {} & L_0\|\bm w_{t+1}-\bar{\bm w}_{t+1}\| \notag   \\
    = {} & L_0\|(\bar{\bm w}_{t+1}-\bm w_t)-(\bm w_{t+1}-\bm w_t)\| \notag \\
    = {} & L_0\|\bm \theta_t-\hat{\bm \theta}^t\|, \label{eqn-1}
\end{align}
where $\bm \theta_t=\bar{\bm w}_{t+1}-\bm w_t$ is the FedAvg aggregation result and $\hat{\bm \theta}^t$ is the PRoBit+ aggregation result with Byzantine attacks and privacy protection. 
Substituting the result of Lemma \ref{lemma-6} back into \eqref{eqn-1}, we get
\begin{align}
    & \mathbb{E}\left[F(\bm w_{t+1})\right]-\mathbb{E}\left[F(\bar{\bm w}_{t+1})\right] \notag \\
    \leq {} & L_0\left(\|\bm \theta_t\|+2\sqrt{2}\beta\|\bm b_t\|+\sqrt{\frac{4\|\bm b_t\|_1}{M}\left(1+\frac{1}{\epsilon}\right)\Delta_1}\right)                \notag                          \\
    \leq {} & \frac{L_0B(1+\gamma)}{\alpha\bar{\lambda}}\|\nabla F(\bm w_t)\|^2 \notag \\
    &+L_0\left(2\sqrt{2}\beta\|\bm b_t\|+\sqrt{\frac{4\|\bm b_t\|_1}{M}\left(1+\frac{1}{\epsilon}\right)\Delta_1}\right). \label{eq3-116}
\end{align}
where \eqref{eq3-116} comes from \eqref{eq3-83} and the rule of stopping the training parameters at the steady-state point, thus the training parameters satisfying $\|\nabla F(\bm w_t)\|> \alpha$.
Combine the above result with Lemma \ref{lemma-5}, we have
\begin{align}
    & \mathbb{E}\left[F(\bm w_{t+1})\right] \notag \\
    \leq {} & \mathbb{E}\left[F(\bm w_{t})\right]-\left(\frac{1-\gamma B}{\lambda}-\left(\frac{L_0}{\alpha}+\frac{L}{\lambda}\right)\frac{B(1+\gamma)}{\bar{\lambda}} \right. \notag \\
    & \left.-\frac{L(1+\gamma)^{2}B^{2}}{2\bar{\lambda}^{2}}\right)\times\left\|\nabla F(\bm w_{t})\right\|^{2} \notag \\
    & +2L_0\left(\sqrt{2}\beta\|\bm b_t\|+\sqrt{\frac{\Delta_1}{M}\left(1+\frac{1}{\epsilon}\right)\|\bm b_t\|_1}\right).
    \label{eq3-118}
\end{align}
\end{myproof}

\section{Proof of Theorem \ref{theorem-3}}\label{theorem-3-proof}

\begin{myproof}[Proof of Theorem \ref{theorem-3}]
    Analyzing the last term of \eqref{eq3-118}, we get
    \begin{align}
        \|\bm b_t\| & = \left\|\bm \theta_t+\bm \zeta\right\|                                                                 \notag     \\
                    & \le \left\|\bm \theta_t\right\|+\left\|\bm\zeta\right\|                                                 \notag    \\
                    & \le \frac{B(1+\gamma)}{\bar{\lambda}}\|\nabla F(\bm w_t)\|+\left\|\bm\zeta\right\|     \label{eq3-123}     \\
                    & \le \frac{B(1+\gamma)}{\bar{\lambda}\alpha}\|\nabla F(\bm w_t)\|^2+\left\|\bm\zeta\right\|, \label{eq3-122}
    \end{align}
    where \eqref{eq3-123} follows from \eqref{eq3-83}, and \eqref{eq3-122} comes from the definition of the steady-state point.
    Therefore, we have
    \begin{align}
        \sqrt{\|\bm b_t\|_1}
         & \le \sqrt{\sqrt{d}\|\bm b_t\|}      \label{eq3-124}                                                                          \\
         & \le \sqrt{\frac{dB(1+\gamma)}{\bar{\lambda}}\|\nabla F(\bm w_t)\|+d\left\|\bm\zeta\right\|}           \notag                        \\
         & \le \sqrt{\frac{dB(1+\gamma)}{\bar{\lambda}}\|\nabla F(\bm w_t)\|}+\sqrt{d\left\|\bm\zeta\right\|}               \notag             \\
         & \le \sqrt{\frac{dB(1+\gamma)}{\bar{\lambda}\alpha^3}}\|\nabla F(\bm w_t)\|^2+\sqrt{d\left\|\bm\zeta\right\|}, \label{eq3-127}
    \end{align}
    where \eqref{eq3-124} follows from the Cauchy-Schwarz inequality. Substituting \eqref{eq3-122} and \eqref{eq3-127} back into \eqref{eq3-118} yields
    \begin{align}
        &\mathbb{E}\left[F(\bm w_{t+1})\right]
        \leq \mathbb{E}\left[F(\bm w_{t})\right]-\rho\left\|\nabla F(\bm w_{t})\right\|^{2} \notag \\
        &+2L_0\left(\sqrt{2}\beta\left\|\bm \zeta\right\|+\sqrt{\frac{\Delta_1}{M}\left(1+\frac{1}{\epsilon}\right)d\left\|\bm\zeta\right\|}\right),
        \label{eq3-129}
    \end{align}
    where
    \begin{align*}
        \rho \triangleq {} & \frac{1-\gamma B}{\lambda}
        -2L_0\sqrt{\frac{\Delta_1}{M}\left(1+\frac{1}{\epsilon}\right)\frac{dB(1+\gamma)}{\bar{\lambda}\alpha^3}} \notag \\
        &-\left(\frac{L_0}{\alpha}+\frac{2\sqrt{2}\beta L_0}{\alpha^2}+\frac{L}{\lambda}\right)\frac{B(1+\gamma)}{\bar{\lambda}} \notag \\
        &-\frac{L(1+\gamma)^{2}B^{2}}{2\bar{\lambda}^{2}} \\
        = {} & \frac{1}{\lambda}-\mathcal{O}\left(B^2\right).
    \end{align*}
    By selecting appropriate $\gamma$, $\lambda$, and $M$ to ensure $\rho>0$, it follows from \eqref{eq3-129} that
    \begin{align*}
        &\frac{1}{T}\sum_{t=1}^{T}\left\|\nabla F(\bm w_{t})\right\|^{2} \le \frac{1}{\rho}\left(\frac{F(\bm w_0)-F^\star}{T}\right. \notag \\
        &+\left. 2L_0\left(\sqrt{2}\beta\left\|\bm \zeta\right\|+\sqrt{\frac{\Delta_1}{M}\left(1+\frac{1}{\epsilon}\right)d\left\|\bm\zeta\right\|}\right)\right).
    \end{align*}
\end{myproof}
}

 





\end{document}